\documentclass[pra,aps,twocolumn,superscriptaddress]{revtex4}
\usepackage{color} 
\usepackage{amssymb} 
\usepackage{graphicx}
\usepackage[english]{babel}
\usepackage{epstopdf}
\usepackage{mathrsfs}
\usepackage{mathtools}
\usepackage{textcomp}
\usepackage{amsfonts}
\usepackage{amssymb} 
\usepackage{hyperref}

\usepackage{quantum}

\begin{document}
\title{Localization transitions and mobility edges in coupled Aubry-Andr\'e chains}
\author{M. Rossignolo}
\affiliation{Dipartimento di Fisica e Astronomia ``G. Galilei'', Universit\`a 
di Padova, I-35131 Padova, Italy}
\affiliation{Institute for Quantum Optics and Center for Integrated Quantum 
Science and Technology, Universit\"at Ulm, D-89081 Ulm, Germany}
\author{L. Dell'Anna}
\affiliation{Dipartimento di Fisica e Astronomia ``G. Galilei'', Universit\`a 
di Padova, I-35131 Padova, Italy}

\begin{abstract}
We study the localization transitions for coupled one-dimensional lattices with quasiperiodic potential. 
Besides the localized and extended phases there is an intermediate mixed phase  which can be easily 
explained decoupling the system so as to deal with effective uncoupled Aubry-Andr\'e chains with different transition points. 
We clarify, therefore, the origin of such an intermediate phase finding the conditions for getting a uniquely defined mobility edge for such coupled systems. 
\textcolor{black}{Finally we consider many coupled chains with an energy shift 
which compose an extension of the Aubry-Andr\'e model in two dimensions.  
We study the localization behavior in this case comparing the results with 
those obtained for a truly aperiodic two-dimensional (2D) Aubry-Andr\'e model, with quasiperiodic potentials in any directions, and for the 2D Anderson model.}
\end{abstract}


\maketitle

 \section{Introduction}
 Since the discovery of the Anderson transition \cite{Anderson1958}, the problem of localization of the wavefunctions in low dimensional 
 quantum systems has attracted a lot of both theoretical and experimental interest in the scientific community 
\cite{Mott,Abrahams,Harper,Aubry,Giamarchi,Billy,Roati}. 
Anderson localization predicts that the single particle wavefunctions become localized in the presence of some uncorrelated disorder, 
leading to a metal-insulator transition caused by the quantum interference in the scattering processes of a particle with random impurities 
and defects. 

Although the standard Anderson transition occurs in three dimensions, an analogous effect may appear in one dimension (1D) 
in the presence of a so-called quasi-disorder. The most popular case is provided by the well celebrated Aubry-Andr\'e model \cite{Harper,Aubry}
which exhibits a transition between a phase where all the eigenstates are localized and another one where they are extended. 
\textcolor{black}{Generalizations of this model 
have been recently proposed by 
exponential short-range hopping \cite{Biddle}, 
flatband networks \cite{flach2}, higher dimensions \cite{r1}, 
power-law hopping \cite{r2}, and breaking the time-reversal symmetry by a magnetic flux \cite{flux}. Dynamical properties of a generalized Aubry-Andr\'e model have also been investigated \cite{r3}.}

Strongly motivated by the recent experiments of Bloch and cowokers \cite{Bordia,Bordia2017,Luschen2017} and by the feasibility of realizing several copies 
of the Aubry-Andr\'e system, 
by cold atoms or optical waveguides \cite{Kraus,Lohse}, 
we investigate the localization transitions for coupled-chains in a quasi-disordered environment. 
Moreover, in the experiment reported in Ref.~\cite{Luschen2017}, 
a coexistence was observed of localized and delocalized states due to the extension of the kinetic term persisting even in  
the strong tight binding limit of a continuous one-dimensional 
bichromatic model, 
although, in that limit, there is not expected to be any range of parameters 
where localized and extended states can coexist. 
\textcolor{black}{
The authors in Ref.~\cite{Luschen2017} explain the discrepancy between the theoretical predictions of a narrower intermediate phase and the observations 
by averaging over many Aubry-Andr\'e chains produced in the experiment, which can have slightly different parameters, due to the finite extension of the beams creating the optical lattices, so that chains on the outside of the system can experience slightly lower lattice depths than those in the center.}
As we will see, 
\textcolor{black}{an analogous effect can be obtained 
if one allows those chains to be coupled. In this case the 
intermediate phase of coexistence may increase and, eventually can 
contribute to the discrepancy discussed above}. 

We will perform, therefore, a systematic study of the localization transitions of two and many coupled chains in order to clarify 
also the appearance of the mobility edges 
in such composed systems. The presence of an intermediate phase where extended and localized states coexist make the definition 
of a mobility edge questionable \cite{Sil2008,Flach2014}. We make clear the origin of such an intermediate phase finding the conditions 
for a unique and well defined mobility edge for such aperiodic coupled chains. 
\textcolor{black}{
We will consider, finally, many coupled chains which have shifted energies 
one compared to the other, obtaining as a result a generalized Aubry-Andr\'e model in two dimensions. We observe that in this case, on average, the extension of the 
wavefunctions, for large quasiperiodic potential is much larger than that 
obtained in the presence of a true uncorrelated disorder. A sharper localization is, instead, obtained by using quasiperiodic potentials in any directions, considering a truly aperiodic 2D Aubry-Andr\'e model \cite{r1}.}

The paper is organized as it follows: in Sec.~II we will briefly review the known results for a single Aubry-Andr\'e model, 
with both nearest-neighbor and further hopping terms, in Sec.~III we will consider two coupled chains for short and longer hopping terms, discussing the intermediate phase, and in Sec.~IV we will generalize the coupling to a generic number of chains. 
\textcolor{black}{Finally, in Sec.~V we will consider the generalization of the Aubry-Andr\'e model in two dimensions, by coupling many chains or imposing aperiodic potentials in both directions, and make the comparison with the Anderson model.} 

\section{The Aubry-Andr\'e model}
\begin{figure}[h!]
\includegraphics[scale=0.24]{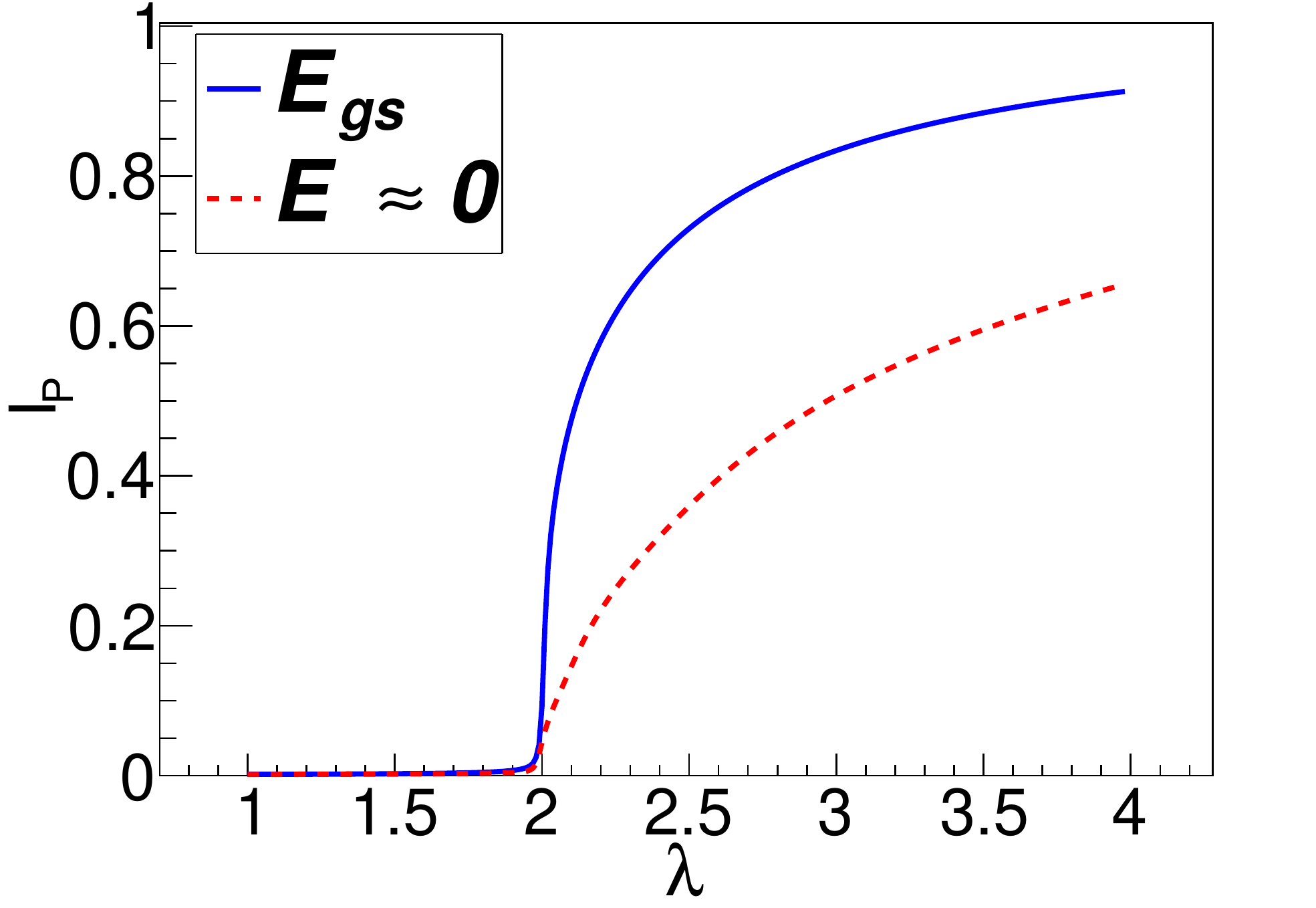}
\caption{IPR for two different eigenvalues of the Aubry-Andr\'e Hamiltonian with nearest-neighbor hopping, with $L=1000$ sites, and $\lambda$ in units of $t_1$.}
\label{fig.IPRstandard}
\end{figure}

Let us consider the following one-dimensional lattice model, called the Aubry-Andr\'e chain, 
\begin{equation}
  \ham = \sum_{i\neq j} t_{ij} \adj{\op{c}}_i \op{c}_{j} + \lambda \sum_{i} \cos(2\pi\tau i)\, \adj{\op{c}}_i \op{c}_{i} 
\end{equation}
where $\tau=\frac{1+\sqrt{5}}{2}$ is the golden ratio, $\adj{\op{c}_{i}}$, 
$\op{c}_{i}$ are the (bosonic or fermionic) creation and annihilation 
operators defined on the lattice site $i$, $t_{ij}$ the hopping parameter and $\lambda$ the strength of the quasi-disordered potential. 
It has been rigorously proven \cite{Jitomirskaya} that 
if the sum is restricted to nearest-neighbor sites, $t_{i j}=t_1(\delta_{j,i+ 1}+\delta_{j,i-1})$, 
 the above system shows a transition at 
\begin{equation}
\lambda=\lambda_c=2 t_1 \,.
 \end{equation}
Above $\lambda_c$ all eigenstates 
are exponentially localized, while below they are all delocalized. 
On the other hand, if the sum is extended to further neighbors,  
there is a mobility edge \cite{Boers}, namely, the critical strength of the potential $\lambda_c$, depends on the energy levels $E_n$, 
and providing that the hopping parameter decays exponentially with the distance ($|t_\ell|\equiv |t_{i,i\pm \ell}|=e^{-p\ell}$ with $p$ some positive real value), the transition can be calculated analytically \cite{Biddle,Ramakumar2014} 
\begin{equation}
\label{lambda_c_extended}
\lambda_c=\frac{2 t_1+2E_n t_2/t_1}{1+(t_2/t_1)^2}\,.
\end{equation}
This expression is exact for exponential form of the hopping parameter but is also in a very good agreement with numerical results if one considers terms up to next-nearest neighbors with $t_2$ much smaller than $t_1$, neglecting further terms which can be assumed 
exponentially small. 
The transition from a localized state to a delocalized one can be detected by the measure of the so called inverse participation ratio (IPR), which is a quantity derived from the eigenfunctions of the hamiltonian, $\ham=\sum_{i,j}\adj{\op{c}}_i H_{ij} \op{c}_j$, 
defined on the lattice with $L$ sites, 
\begin{equation}
\label{eq.AA}
\sum_{j=1}^L H_{ij}\psi_{n,j}=E_n \psi_{n,i}
\end{equation} 
so that the IPR is defined for any eigenstate
\begin{equation}
\label{IPRn}
I_{P}^{(n)}=\frac{\sum_i |\psi_{n,i}|^4}{\sum_i |\psi_{n,i}|^2}\,. 
\end{equation}
For normalized wavefunctions, $\sum_i |\psi_{n,i}|^2=1$, one gets $0\le I_P^{(n)}\le 1$. The two extreme limits can be explain as it follows. 
For a very extended state $|\psi_{n,i}|\sim 1/\sqrt{L}$, therefore $I_P^{(n)}\sim 1/L$, which goes to zero in the thermodynamic limit, 
while for a strongly localized state $|\psi_{n,i}|\sim \delta_{i, i_0}$, so that $I_P^{(n)}\sim 1$. In Fig.~\ref{fig.IPRstandard}, as examples, 
the IPRs of two eigenstates are reported, related to the ground state and to a state at the band energy center for the Aubry-Andr\'e model with nearest-neighbor hopping. The value of $\lambda$ for which the IPR drops to zero is the critical point $\lambda_c$.
 \section{Two coupled chains}
Let us now consider two copies of the Aubry-Andr\'e chain coupled together by some additional transverse hopping parameters. 
The general form of the Hamiltonian is the following
\begin{align}
 \ham = \sum_{i \neq j} \left(t_{ij} \adj{\op{c}}_{i} \op{c}_{j} + t_{ij} \adj{\op{d}}_{i} \op{d}_{j} \right)+
\sum_{i \neq j} t^{d}_{ij} \left(\adj{\op{c}}_{i} \op{d}_{j} + \adj{\op{d}}_{j} \op{c}_{i}\right)   \nonumber \\ 
 + \sum_{i} t^{d}_0 \left(\adj{\op{c}}_{i} \op{d}_{i} + \adj{\op{d}}_{i} \op{c}_{i}\right) + 
 \sum_{i} \epsilon(i)\left(\adj{\op{c}}_{i} \op{c}_{i} + \adj{\op{d}}_{i} \op{d}_{i} \right) 
 \label{eq:ham1}
\end{align}
where $\epsilon(i)=\lambda \cos(2 \pi \tau i)$ are the on-site energies, $t_{ij}$ the hopping parameter between sites of the same chain, 
$t^{d}_{ij}$ and $t^{d}_0$ are the hopping parameters between sites belonging to different chains and with different or same 
on-site energies respectively, $\op{c}$ and $\op{d}$ are the operators defined on the two different chains. 
In the following subsections we will study numerically and analytically the localization transitions for this system.
\subsection{Nearest-neighbor hopping}
\begin{figure}[h] 
  \centering
\includegraphics[width=0.4\textwidth]{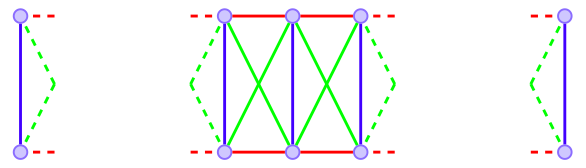}
 \caption{Two coupled chains: the red line describes the nearest-neighbor intra-chain hopping $t_{1}$, 
the blue line the transverse hopping $t^d_{0}$, the green line the nearest-neighbor inter-chain hopping $t^d_{1}$.}
\label{figNNA}
 \end{figure}

We first review the system of two chains coupled by nearest-neighbor hopping, introduced in Ref.~\cite{Sil2008} and 
commented on in Ref.~\cite{Flach2014}. 
This model is described by Fig.~\ref{figNNA}. 

\begin{figure}[th!]
  \centering
  \includegraphics[scale=0.24]{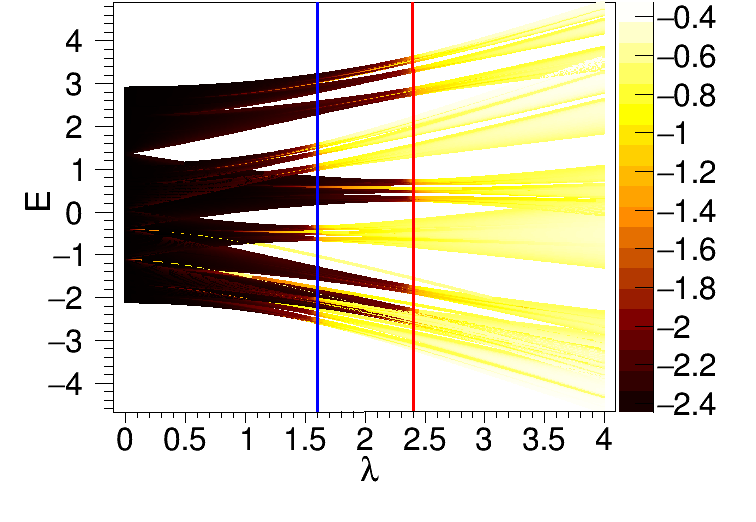}
  \qquad\qquad
  \includegraphics[scale=0.32]{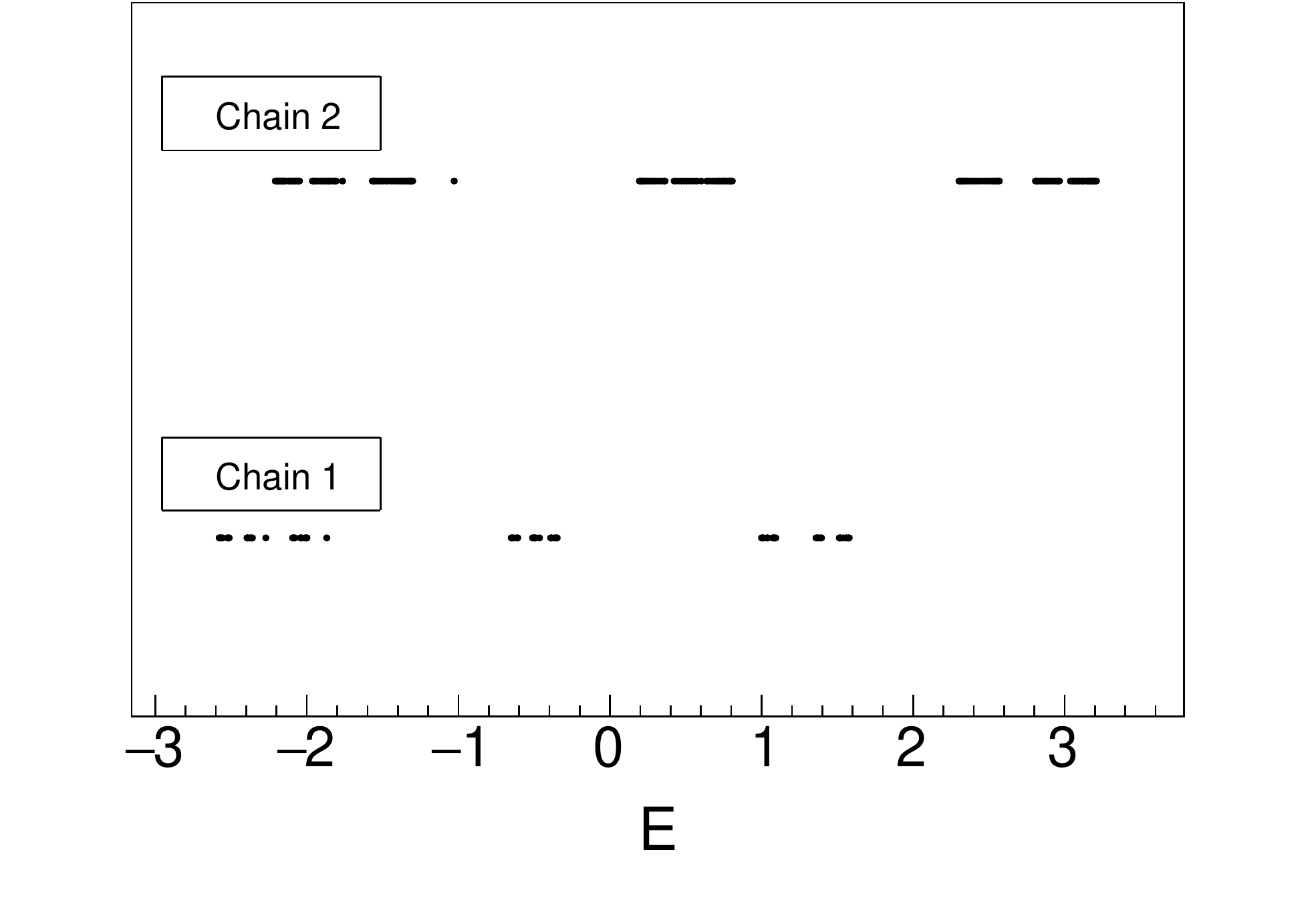}
  \caption{Inverse participation ratio (IPR), in \textcolor{black}{$10$-base} logarithmic scale, for two \textcolor{black}{coupled} Aubry-Andr\'e chains \textcolor{black}{with $L=200$ sites each and} with nearest-neighbor hopping $t_{1} = 1$, $t^d_{1} = 0.2$ and $t^d_{0} = 0.5$. \textcolor{black}{The two vertical lines are given by Eq.~(\ref{eq.lambda+}) (red right line) and Eq.~(\ref{eq.lambda-}) (blue left line).} (Below) Energy spectra, $E^+$ and $E^-$, of the two effective decoupled chains at $\lambda = \lambda_c^-=1.6$.}
  \label{fig:LLs200t11t21td02tds05}
\end{figure}

In terms of the spinor
\begin{equation}
 \op{b} =\left(
 \begin{matrix}
  \op{c} \\
  \op{d}
 \end{matrix}\right)\,,
\end{equation}
the Hamiltonian can be written
\begin{align}
\label{Hspinor}
\ham = \sum_{i} \adj{\op{b}}_{i} {\cal{E}}(i) \op{b}_{i} + \sum_{i} \left(\adj{\op{b}}_{i} {T_1}\, \op{b}_{i+1}
+ \adj{\op{b}}_{i+1} {T_1}\, \op{b}_{i}\right) ,
\end{align}
where
\begin{equation}
{\cal E}(i) =\left(
\begin{matrix}
 \epsilon(i) & t^d_{0}      \\
 t^d_{0}       & \epsilon(i)
\end{matrix}\right)
\end{equation}
and
\begin{equation}
T_1 =\left(
\begin{matrix}
 t_{1} & t^d_{1}      \\
 t^d_{1}       & t_{1}
\end{matrix}\right) ,
\end{equation}
where
$t_{1}=t_{i,i\pm 1}$ is 
the nearest-neighbor hopping between the $i$-th and the $(i\pm 1)$-th site of the same chain, $t^d_{0}$ is the transverse nearest-neighbor 
inter-chain hopping between the two chains and $t^d_{1}=t^d_{i,i\pm 1}$ is the nearest-neighbor inter-chain 
hopping between the $i$-th and the $(i\pm 1)$-th site of the two different chains 
(actually, it is already a next-nearest-neighbor hopping parameter between the two neighboring chains).  
Introducing the wavefunction
\begin{equation}
 \Psi_{n,i} = \left(
 \begin{matrix}
  \psi^{(1)}_{n,i}\\
  \psi^{(2)}_{n,i}
 \end{matrix}\right)
\end{equation}
where $\psi^{(\alpha)}_{n,i}$ are the amplitudes of the wavefunctions at the $i$-th site of the $\alpha$-th chain ($\alpha = 1,2$). 
The Schr\"odinger equation in this basis, can be written as 
\begin{equation}
 \left( E_n \mathbb{1} - {\cal E}(i) \right){\Psi}_{n,i} = T_1\left( \Psi_{n,i+1} + \Psi_{n,i-1} \right)
\end{equation}
which explicitly corresponds to the following coupled equations
\begin{eqnarray}
\nonumber \big(E_n - \epsilon(i) \big)\psi^{(1)}_{n,i} - t^d_{0}\psi^{(2)}_{n,i}  &=& t_{1}\big( \psi^{(1)}_{n,i+1} + \psi^{(1)}_{n,i-1} 
\big)  \\
&+& t^d_{1}\big( \psi^{(2)}_{n,i+1} + \psi^{(2)}_{n,i-1} \big)\\
\nonumber \big(E_n - \epsilon(i) \big)\psi^{(2)}_{n,i} - t^d_{0}\psi^{(1)}_{n,i}  &=& t_{1}\big(\psi^{(2)}_{n,i+1} + \psi^{(2)}_{n,i-1} 
\big)  \\
&+& t^d_{1}\big( \psi^{(1)}_{n,i+1} + \psi^{(1)}_{n,i-1} \big)
\end{eqnarray}
Applying the following canonical transformation
\begin{equation}
\label{transform}
\psi_{n,i}^{\pm} = \frac{\psi^{(1)}_{n,i} \pm \psi^{(2)}_{n,i}}{\sqrt{2}}
\end{equation}
the system is exactly mapped to two uncoupled Aubry-Andr\'e chains described by Eq.~(\ref{eq.AA}), explicitly,
\begin{eqnarray}
\label{eq:1chainNN}
&& \hspace{-1cm} [E_n - (\epsilon(i) - t^d_{0}) ] \psi_{n,i}^{-} = (t_{1} - t^d_{1})( \psi_{n,i+1}^{-} + \psi_{n,i-1}^{-} ) , \\
&& \hspace{-1cm} [E_n - (\epsilon(i) + t^d_{0}) ] \psi_{n,i}^{+} = (t_{1} + t^d_{1})( \psi_{n,i+1}^{+} + \psi_{n,i-1}^{+} ) .
\label{eq:2chainNN}
\end{eqnarray}
The full spectrum $E_m$ is composed by two different spectra 
$E_n^{-}=E_n+t^d_0$ and $E_n^{+}=E_n-t^d_0$ 
of two uncoupled Aubry-Andrè chains whose localization transitions 
occur at 
\begin{eqnarray}
\label{eq.lambda+}
&&\lambda^+_{c} = 2(t_{1} + t^d_{1}) \\ 
&&\lambda^-_{c} = 2(t_{1} - t^d_{1})
\label{eq.lambda-}
\end{eqnarray}
and can be sorted in ascending order labeling $m=1,\dots,2L$, so that, for $t^d_0>0$, $E_0-t^d_0\le E_m\le E_L+t^d_0$. The corresponding 
eigenstates $\psi_{m,i}$ are equal to some $\psi^+_{n,i}$ or $\psi^-_{n,i}$ depending on the energy level.   
In particular, for $\lambda$ between the two critical values, the eigenstates $\psi_m$ 
are localized or delocalized depending on whether they correspond to $\psi^-_n$ or $\psi^+_n$.   
This argument can explain Fig.~\ref{fig:LLs200t11t21td02tds05} where the 
IPR, $I_P^{(m)}=\sum_i|\psi_{m,i}|^4/\sum_i|\psi_{m,i}|^2$, is shown, in logarithmic scale, 
as a function of $\lambda$ and energy. Below $\lambda_c^-$ all states are delocalized, above $\lambda_c^+$ are all localized and in between, in an intermediate phase, they coexist. In Fig.~\ref{fig:LLs200t11t21td02tds05}, in the plot below, the two spectra $E^+$ and $E_-$ of the two effective chains are reported for a specific value of $\lambda$. Since there is a shift of $2t^d_0$ the nature of the states at the external bands is dictated only by one of the two effective Aubry-Andr\'e chains. 
In this situation it is questionable speaking about the presence of a mobility edge, 
as explained in Ref.~\cite{Flach2014}. 
As final remark, one can notice that for the pure ladder configuration, with $t^d_1=0$, the two effective uncoupled 
chains differ only by an energy shift while the critical values are the same as they are without any coupling. 

\subsection{Generalization to longer range hopping and the case with next-nearest neighbors}
We can generalize the previous results to many-neighbor hopping parameter. The Hamiltonian can be written as:
\begin{align}
\label{Hami}
\ham = \sum_{i} \adj{\op{b}}_{i} {\cal E}(i) \op{b}_{i} + \sum_{i,\ell} \left( \adj{\op{b}}_{i} T_\ell \op{b}_{i+\ell}
+ \adj{\op{b}}_{i+\ell} T_\ell \op{b}_{i}\right)
\end{align}
where now $T_\ell$ is defined by
\begin{equation}
T_\ell =\left(
\begin{matrix}
 t_{\ell} & t^d_{\ell}      \\
 t^d_{\ell}       & t_{\ell}
\end{matrix}\right)
\end{equation}
and $t_{\ell}=t_{i,i\pm \ell}$, $t^d_{\ell}=t^d_{i,i\pm \ell}$ are the $\ell$-th-neighbor hopping terms for the sites 
belonging to the same chain and to different chains respectively. 
Introducing the wavefunction and by the transformation (\ref{transform}), in the same way as before, one gets two decoupled 
Schr\"odinger equations
\begin{eqnarray}
\label{general1}
\big[ E_n - (\epsilon(i) + t^d_{0}) \big]\psi_{n,i}^{+} =
\sum_{\ell}(t_{\ell} + t^d_{\ell})( \psi_{n,i+\ell}^{+} + \psi_{n,i-\ell}^{+} )\;\; \\
  \big[ E_n - (\epsilon(i) - t^d_{0}) \big]\psi_{n,i}^{-} =
\sum_{\ell}(t_{\ell} - t^d_{\ell})( \psi_{n,i+\ell}^{-} + \psi_{n,i-\ell}^{-} ) \;\;
\label{general2}
\end{eqnarray}
describing two uncoupled extended Aubry-Andr\'e models. 

\subsubsection{Next-nearest-neighbor hopping}
\begin{figure}[h]
\includegraphics[width=0.4\textwidth]{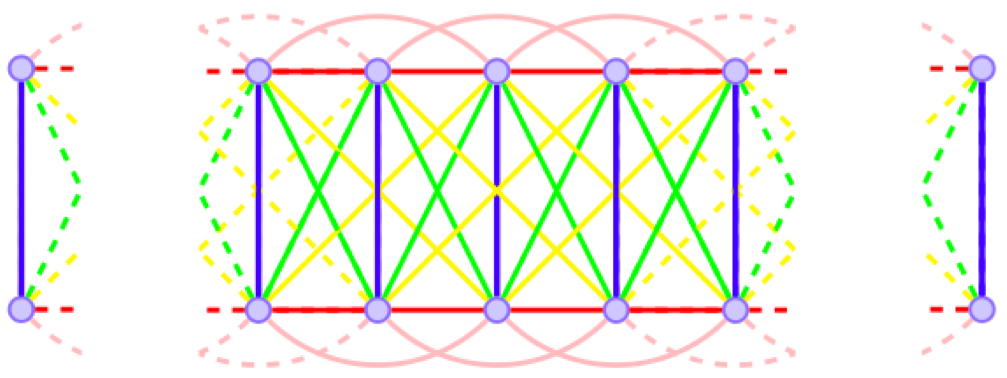}
 \caption{Two coupled chains: the red line describes the nearest-neighbor hopping $t_{1}$, the pink line the second-nearest-neighbor hopping $t_{2}$, the blue line the transverse hopping $t^d_{0}$, the
 green line the nearest-neighbor inter-chain hopping $t^d_{1}$, the yellow line the second-nearest-neighbor inter-chain hopping $t^d_{2}$.}
\label{fig.NNN}
 \end{figure}

\begin{figure}[ht]
  \centering
  \includegraphics[scale=0.24]{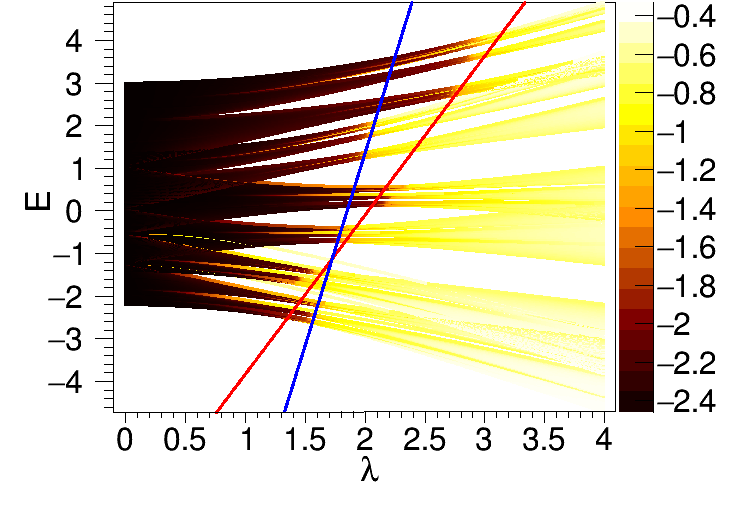}
  \qquad\qquad
  \includegraphics[scale=0.32]{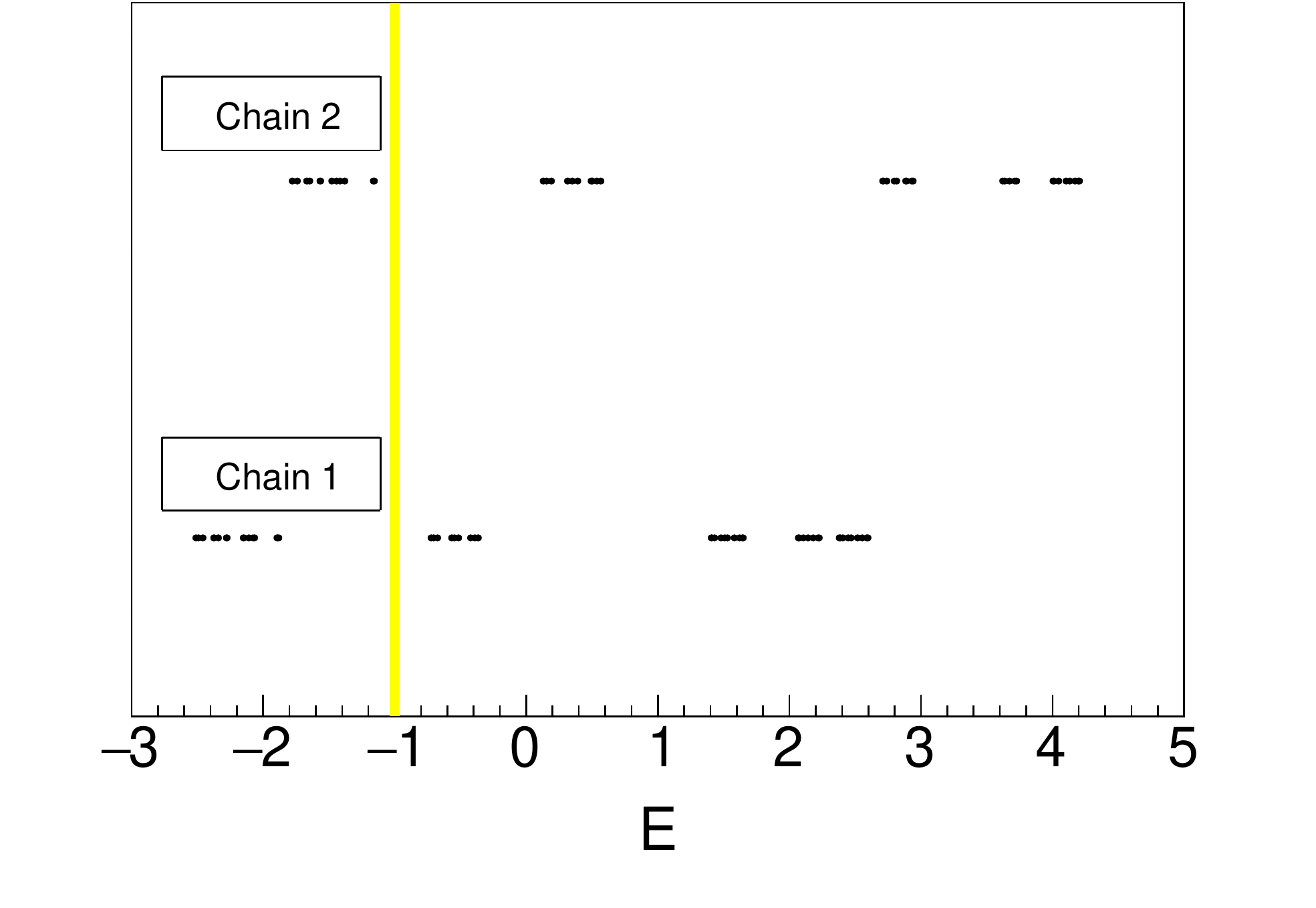}
  \qquad\qquad
  \includegraphics[scale=0.32]{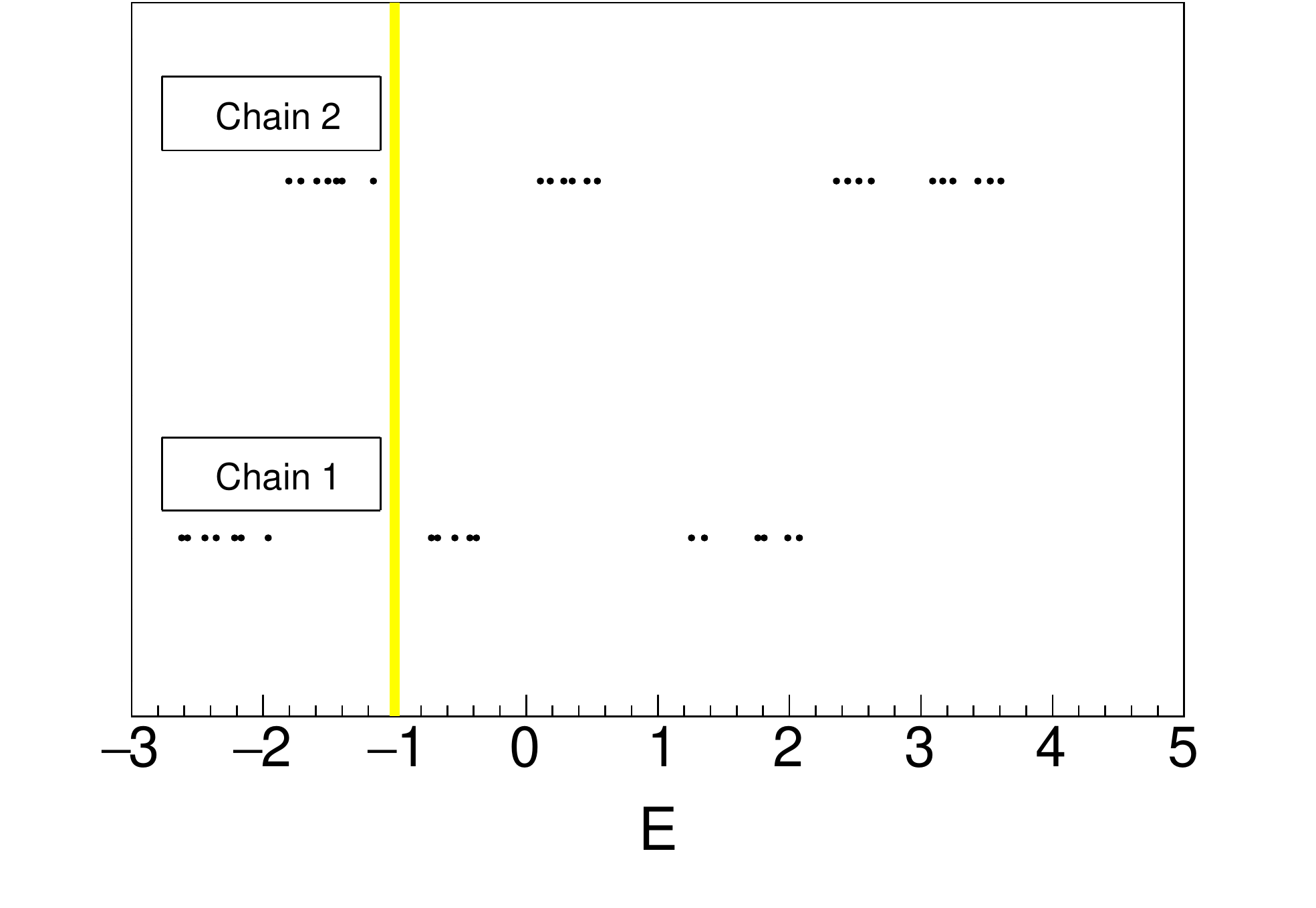}
  \caption{Inverse participation ratio (IPR), in \textcolor{black}{$10$-base} logarithmic scale, 
for two \textcolor{black}{coupled} Aubry-Andr\'e chains \textcolor{black}{($L=200$)} with second-nearest-neighbor hopping: 
$t_{1} = 1$, $t^d_{1} = 0.1$, $t_{2} = 0.1$, $t^d_{2} = 0.05$ and $t^d_{0} = 0.5$. \textcolor{black}{The lines are given by Eq.~(\ref{eq:1lambdaNNN}) (blue line) and Eq.~(\ref{eq:2lambdaNNN}) (red line) in clockwise order.}
(Below) Energy spectra, $E^+$ and $E^-$, of the two effective decoupled chains along $\lambda=\lambda^+_{c}$ \textcolor{black}{(upper plot)} and $\lambda=\lambda^-_{c}$ \textcolor{black}{(lower plot)}. The yellow vertical line indicates the crossing energy point $E^*$.}
  \label{fig:LLs200t11t1201t21t2201td01td2005tds05}
\end{figure}

Let us consider the case with second-nearest-neighbor hopping, that can be described in Fig.~\ref{fig.NNN} 
so that Eqs.~(\ref{general1}), (\ref{general2}) becomes simply
\begin{eqnarray}
\nonumber \big[ E_n - (\epsilon(i) - t^d_{0}) \big] \psi_{n,i}^{-} = (t_{1} - t^d_{1})( \psi_{n,i+1}^{-} + \psi_{n,i-1}^{-} ) \\
+(t_{2} - t^d_{2})( \psi_{n,i+2}^{-} + \psi_{n,i-2}^{-} )\label{eq:1chainNNN}\\
\nonumber \big[ E_n - (\epsilon(i) + t^d_{0}) \big] \psi_{n,i}^{+} = (t_{1} + t^d_{1})( \psi_{n,i+1}^{+} + \psi_{n,i-1}^{+} ) \\
+(t_{2} + t^d_{2})( \psi_{n,i+2}^{+} + \psi_{n,i-2}^{+} )\label{eq:2chainNNN}
\end{eqnarray}
which are two uncoupled chains. 
In the hypothesis of $(t_2\pm t^d_2)\ll (t_1\pm t^d_1)$ and neglecting further terms which can be assumed exponentially small, 
we can resort to the analytical result reported in Eq.~(\ref{lambda_c_extended}) for a single extended Aubry-Andr\'e 
model \cite{Biddle,Ramakumar2014}, getting the 
following values of the critical potentials
\begin{eqnarray}
    \lambda^-_{c} = \frac{2 (t_{1} - t^d_{1}) + 
2(E_n+t^d_{0})\left(\frac{t_{2} - t^d_{2}}{t_{1} - t^d_{1}}\right)}
{1 + \left(\frac{t_{2} - t^d_{2}}{t_{1} - t^d_{1}}\right)^2}
\label{eq:1lambdaNNN} \\
    \lambda^+_{c} = \frac{2 (t_{1} + t^d_{1}) + 
2(E_n-t^d_{0})\left(\frac{t_{2} + t^d_{2}}{t_{1} + t^d_{1}}\right)}
{1 + \left(\frac{t_{2} + t^d_{2}}{t_{1} + t^d_{1}}\right)^2}
\label{eq:2lambdaNNN}
\end{eqnarray}
As one can see from Fig.~\ref{fig:LLs200t11t1201t21t2201td01td2005tds05}, the full spectrum results from the overlap of the spectra 
of two uncoupled chains so that in general one can have an intermediate phase defined as a regime of parameters for which we 
have a coexistence of localized and delocalized states. 

\begin{figure}
\includegraphics[scale=0.24]{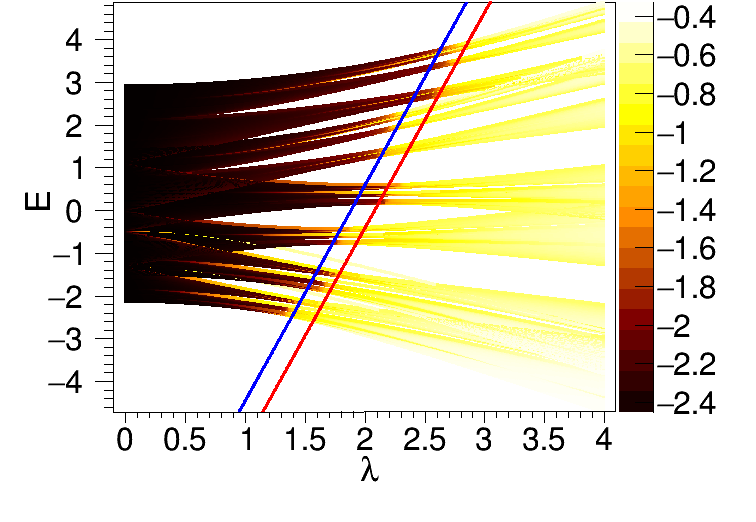}
\qquad\qquad
\includegraphics[scale=0.32]{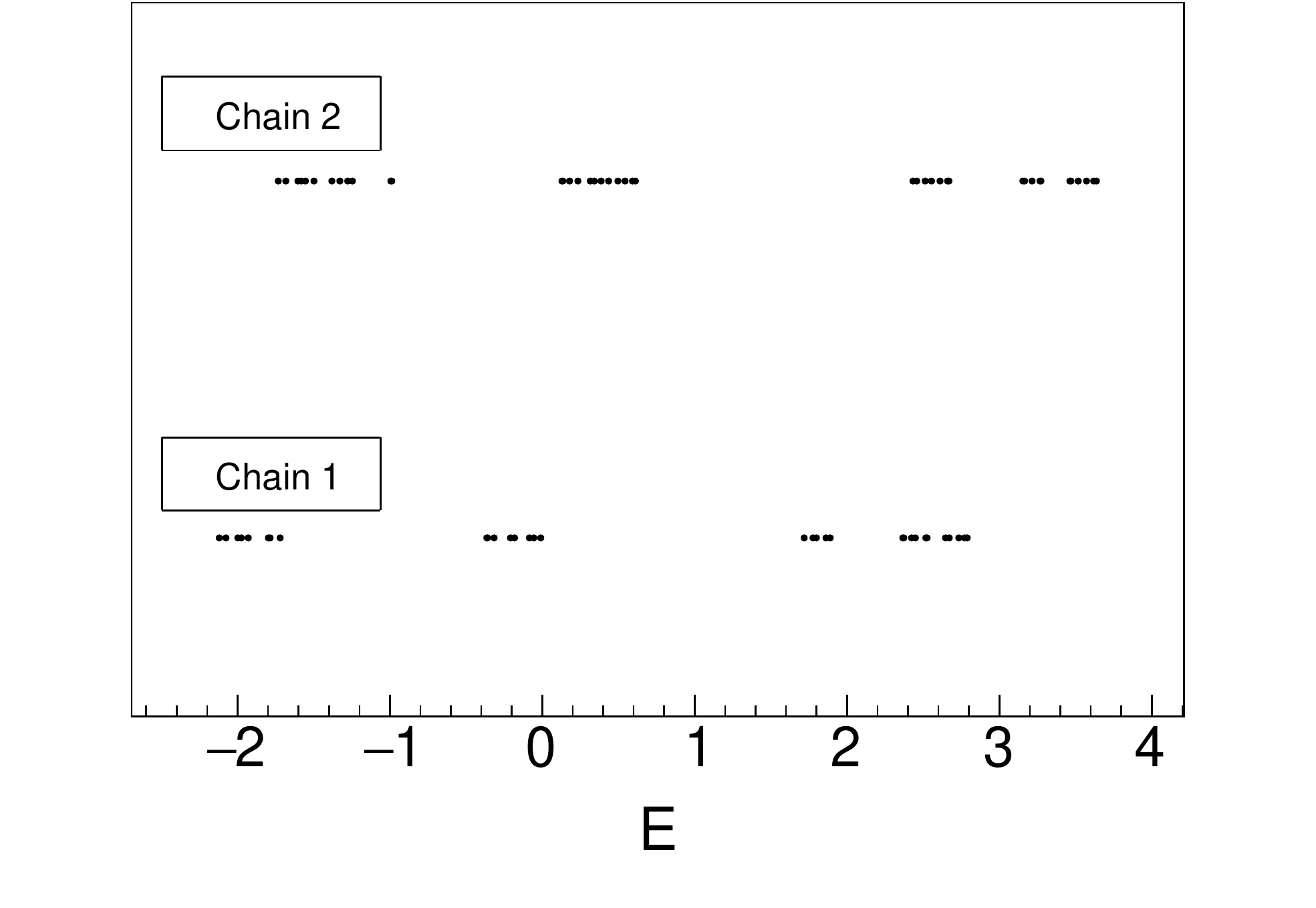}
  \caption{Inverse participation ratio (IPR), in \textcolor{black}{$10$-base} logarithmic scale, for two \textcolor{black}{coupled} Aubry-Andr\'e chains \textcolor{black}{($L=200$)} with second-nearest-neighbor hopping:
$t_{1} = 1$, $t^d_{1} = 0.1$, $t_{2} = 0.1$, $t^d_{2} = 0.01$ and $t^d_{0} = 0.5$. 
\textcolor{black}{The lines are described by Eq.~(\ref{eq:1lambdaNNN}) 
(blue left line) and Eq.~(\ref{eq:2lambdaNNN}) (red right line).} 
(Below) Energy spectra, $E^+$ and $E^-$, of the two effective decoupled chains along $\lambda=\lambda^-_{c}$.}
  \label{fig:LLs200t11t1201t21t2201td01td2001tds05}
\end{figure}

\begin{figure}[h!]
  \centering
  \includegraphics[scale=0.24]{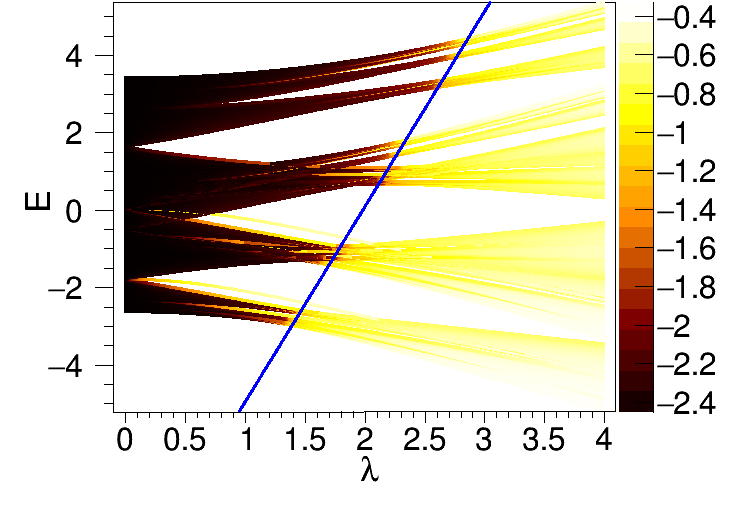}
  \qquad\qquad
  \includegraphics[scale=0.32]{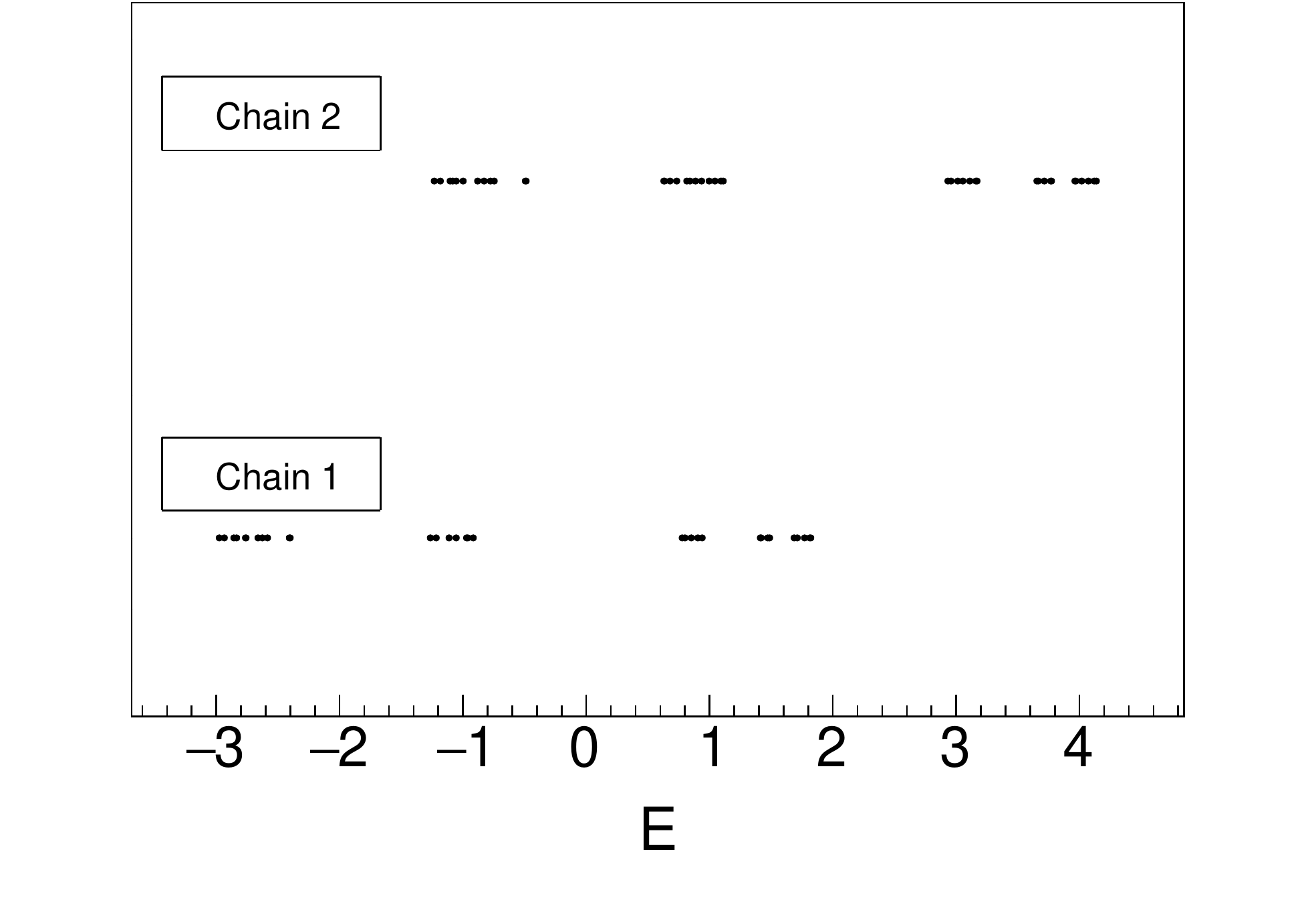}
  \caption{
Inverse participation ratio (IPR), in \textcolor{black}{$10$-base} logarithmic scale, for two \textcolor{black}{coupled} Aubry-Andr\'e chains 
\textcolor{black}{($L=200$)} with second-nearest-neighbor hopping: 
$t_{1} = 1$, $t^d_{1} = 0.1$, $t_{2} = 0.1$, $t^d_{2} = 0.01$ and $t^d_{0} = 1$. \textcolor{black}{Since Eqs.~(\ref{cond1}), (\ref{cond2}) are fulfilled, 
so that the critical line can be written in terms of only the ratio $t_1/t_2$, the mobility edge, described by the blue straight line, is given 
by Eq.~(\ref{lambda_c_extended}).}
(Below) Energy spectra, $E^+$ and $E^-$, of the two effective decoupled chains along $\lambda=\lambda^-_{c}=\lambda^+_{c}$.}
  \label{fig:LLs200t11t1201t21t2201td01td2001tds1}
\end{figure}

For the sake of simplicity of notation let us define
\begin{eqnarray}
\label{A+}
 A_+
=\frac{t_{2} + t^d_{2}}{t_{1} + t^d_{1}}\\
 A_-
= \frac{t_{2} - t^d_{2}}{t_{1} - t^d_{1}}
\label{A-}
\end{eqnarray}
As shown in Fig.~\ref{fig:LLs200t11t1201t21t2201td01td2005tds05}, the localized and delocalized states are delimited by the transition 
lines defined by Eqs.~(\ref{eq:1lambdaNNN}),~(\ref{eq:2lambdaNNN}) which are straight lines as functions of the energy  
with slopes $2A_\pm/(1+A_{\pm}^2)$. In order to determine the crossing point one can impose 
the condition $\lambda^+_{c} = \lambda^-_{c}$ and solve the equation for the energy, getting
\begin{eqnarray}
\nonumber &&\hspace{-0.45cm} E^{*}  = \frac{1}{(A_+ - A_-)(1- A_+A_-)}\Big[ t^d_{0}(1+ A_+A_-)(A_+ + A_-) \\
&&\phantom{E^{*} =} - t^d_{1}(2+A_+^{2} + A_-^{2})+ t_{1}(A_+^{2} - A_-^{2})\Big] 
\end{eqnarray}
for $A_+\neq  A_-$ and $A_+ \neq 1/A_- $, 
as shown in Fig.~\ref{fig:LLs200t11t1201t21t2201td01td2005tds05}. On the other hand, if $A_+= A_-$ (or $A_+ = 1/A_-$), the two lines are parallel as in the case of Fig.~\ref{fig:LLs200t11t1201t21t2201td01td2001tds1}. The most relevant physical situation is when $A_+= A_-$ \textcolor{black}{(in agreement with the 
hypothesis $t_{2} \pm t_{2}^{d} \ll t_{1} \pm t_{1}^{d}$)},    
namely when
\begin{equation}
\label{cond1}
\frac{t_{1}}{t_2} = \frac{t^d_{1}}{t^d_{2}} .
\end{equation} 
By this condition we have two parallel critical lines with slope $\frac{2(t_2/t_1)}{1 + \left( t_{2}/t_{1}\right)^{2}}$ 
and a coexisting region where there are localized and delocalized states, as show 
in Fig.~\ref{fig:LLs200t11t1201t21t2201td01td2001tds05}. 
If we now impose an additional condition to Eq.~(\ref{cond1}) for the hopping parameters
\begin{equation}
\label{cond2}
\frac{t^d_{0}}{t^d_{1}} = \frac{t_{1}}{t_{2}}
\end{equation}
which is also quite reasonable, we get that $\lambda^+_c=\lambda^-_c$ for any value of the energy. \textcolor{black} {From Eqs.~(\ref{cond1}) and (\ref{cond2})
we can express the critical potential in terms of only $t_1/t_2$, getting 
the same result as that of a single chain with exponentially short-range hopping, 
Eq.~(\ref{lambda_c_extended}).}
This means that we get a uniquely 
defined mobility edge which separates the localized phase from the delocalized 
one, as shown in  
Fig.~\ref{fig:LLs200t11t1201t21t2201td01td2001tds1}. The above results, to our knowledge, have not been presented before.

\subsection{Intermediate phase: coexistence of extended and localized states}
\begin{figure}[h!]
  \centering
  \includegraphics[scale=0.32]{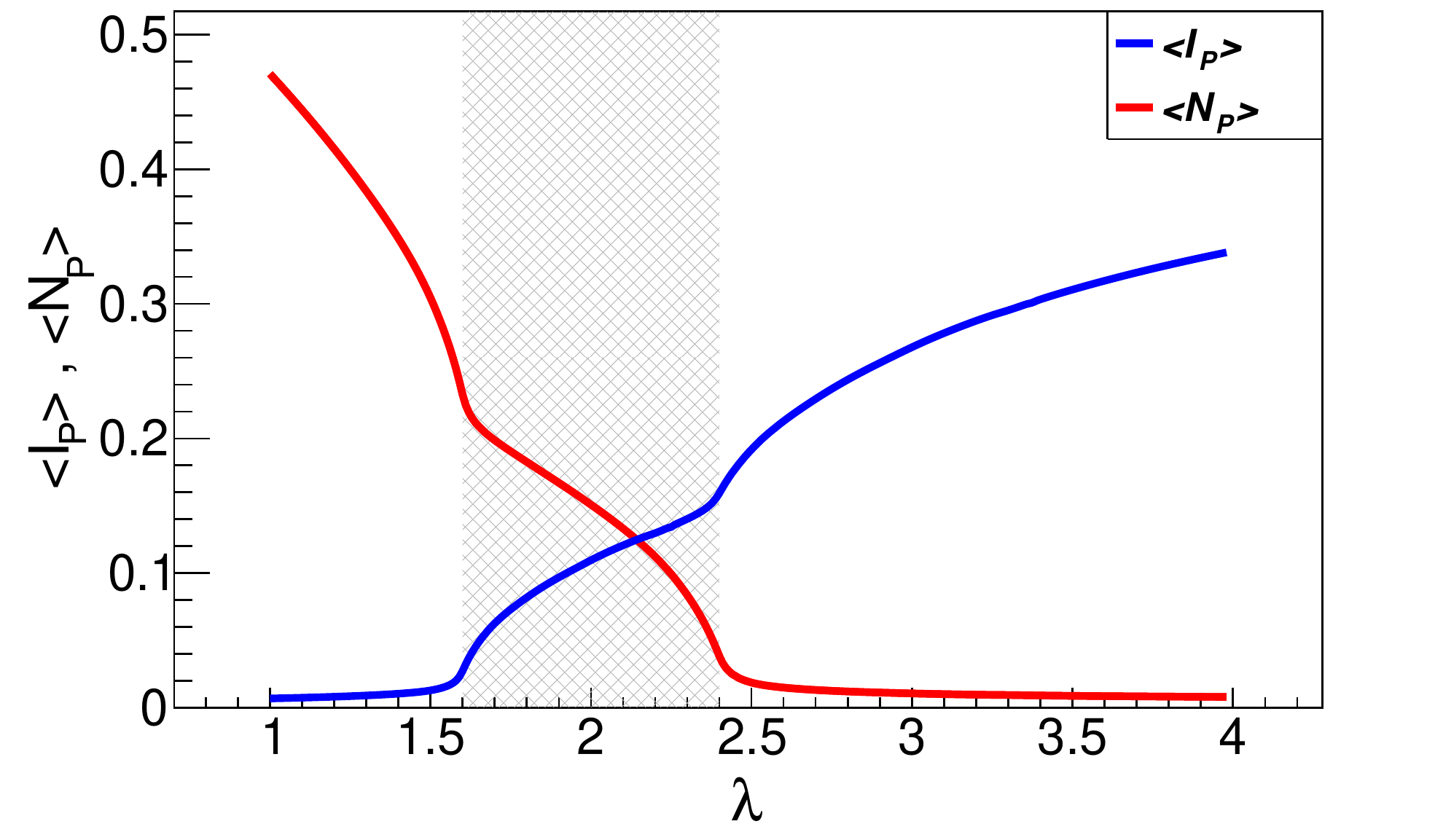}
  \qquad\qquad
  \includegraphics[scale=0.32]{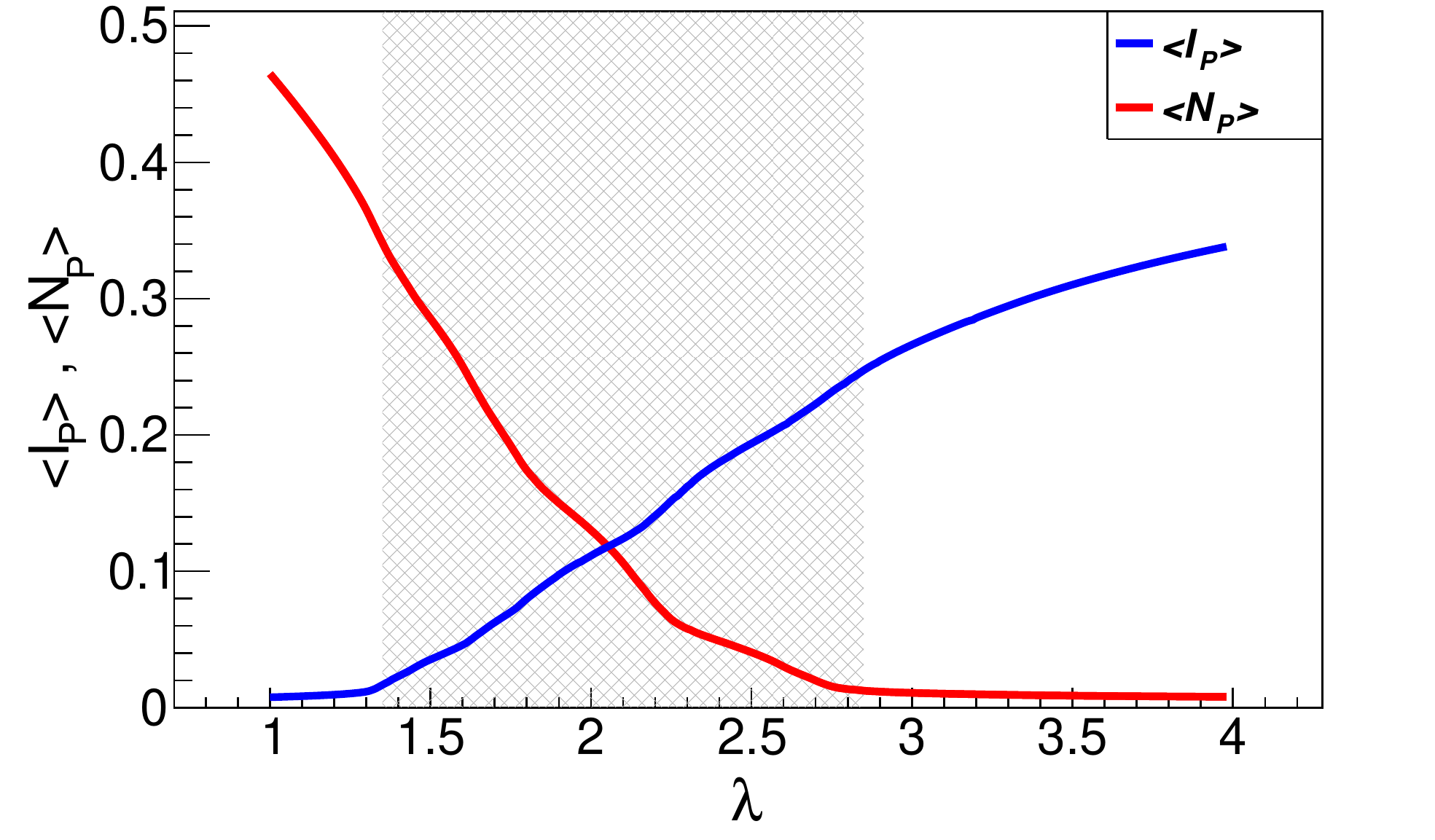}
  \caption{The averaged IPR and NPR, $\medio{I_P}$ and $\medio{N_P}$, obtained from the spectrum reported in Fig.~\ref{fig:LLs200t11t21td02tds05} 
(nearest-neighbor case) and in Fig.~\ref{fig:LLs200t11t1201t21t2201td01td2001tds05} (next-nearest-neighbor case).} 
  \label{fig:iprnprmeanv}
\end{figure}

We can define two different quantities that draw the contour of the region of parameters where localized and 
delocalized states coexist \cite{Li2017}.\\
From the definition of \textcolor{black}{IPR} for an arbitrary state, Eq.~(\ref{IPRn}),  
we can take the average over a set of energy levels whose number is $N_L$ 
\begin{equation}
 \medio{I_P} = \sum_{n}^{N_L} \frac{I_P^{(n)}}{N_L}
\end{equation}
which vanishes when all the $N_L$ states are extended. 
One can use also a complementary quantity, the normalized participation ratio (NPR) 
\begin{equation}
 N_{P}^{(n)} = \frac{1}{L \sum_{i} |\psi_{n,i}|^{4}}
\end{equation}
and, analogously, from that, one defines its average over a subset of states, 
\begin{equation}
 \medio{N_P} = \sum_{n}^{N_L} \frac{N_P^{(n)}}{N_L}
\end{equation}
where $L$ is the size of the system, which vanishes when all the $N_L$ states are localized. \\
In the regime where both $\medio{I_P}$ and $\medio{N_P}$ remain finite, the spectrum of the Hamiltonian 
allows for a phase which has both spatially extended and localized eigenstates. 
This behavior defines an intermediate phase (the shaded regions in Fig.~\ref{fig:iprnprmeanv}) 
made by a mixture of extended and localized states. 
In Fig.~\ref{fig:iprnprmeanv} we plotted $\medio{I_P}$ and $\medio{N_P}$
got from averaging over all the eigenstates, for nearest-neighbor and next-nearest-neighbor cases of two coupled Aubry-Andr\'e models.
This intermediate phase can be detected also dynamically, as shown in Refs.~\cite{r3,Luschen2017},
measuring a finite density imbalance between even and odd sites due to an initial charge density wave state persisting in time.

\section{Many coupled chains: square lattice with quasi-periodicity in one direction}
Let us now consider a further generalization, 
coupling $S$ identical Aubry-Andr\'e chains, as in Ref.~\cite{Guo}, so that the system is described by the following Hamiltonian
\begin{align}
 \ham = \sum_{i \neq j, \alpha} t_{ij} \adj{\op{c}}_{i,\alpha} \op{c}_{j,\alpha} + \textcolor{black}{\frac{1}{2}}
\sum_{i \neq j, \alpha \neq \beta} t^{d}_{ij,\alpha\beta} \left(\adj{\op{c}}_{i,\alpha} \op{c}_{j,\beta} + \adj{\op{c}}_{j,\beta} \op{c}_{i,\alpha}\right)  \nonumber \\ 
 + \textcolor{black}{\frac{1}{2}}
\sum_{i,\alpha \neq \beta} t^d_{0,\alpha\beta} \left(\adj{\op{c}}_{i,\alpha} \op{c}_{i,\beta} + \adj{\op{c}}_{i,\beta} \op{c}_{i,\alpha}\right) + 
\sum_{i,\alpha} \epsilon(i)
\adj{\op{c}}_{i,\alpha} \op{c}_{i,\alpha} 
 \label{eq:ham1}
\end{align}
For simplicity, we will consider only couplings between nearest-neighbor chains, namely $\sum_{\alpha\neq \beta}$ represents $\sum_{\langle \alpha, \beta\rangle}$, a sum over nearest-neighbor chains, so that we can keep using the same notation as before, 
$t^d_0 \equiv t^d_{0,\alpha \,\alpha\pm 1}$ and $t^d_{ij} \equiv t^d_{ij,\alpha\,\alpha\pm 1}$, where $t^d_{i\, i\pm \ell}=t^d_\ell$. 
By these definitions, the Hamiltonian can be \textcolor{black}{rewritten} as in Eq.~(\ref{Hami})
where now 
\begin{equation}
 \op{b}_{i} =\left( 
 \begin{matrix}
  \op{c}_{i,1} \\
  \vdots \\
  \op{c}_{i,S}
 \end{matrix}\right)
\end{equation}
and the $S\times S$ matrices
\begin{equation}
\label{calES}
{\cal E}(i)=\left(
\begin{matrix}
  \epsilon(i) & t^d_{0}            & 0            & \dots  & 0            \\
  t^d_{0}            & \epsilon(i) & t^d_{0}            & \ddots & \vdots       \\
  0            & t^d_{0}            & \epsilon(i) & \ddots & 0            \\
  \vdots       & \ddots       & \ddots       & \ddots & t^d_{0}            \\
  0            & \dots        & 0            & t^d_{0}      & \epsilon(i) \\
  \end{matrix}\right)
\end{equation}
and
\begin{equation}
\label{TjS}
T_\ell=\left(
\begin{matrix}
  t_{\ell} & t^d_{\ell}            & 0            & \dots  & 0            \\
  t^d_{\ell}            & t_{\ell} & t^d_{\ell}            & \ddots & \vdots       \\
  0            & t^d_{\ell}            & t_{\ell} & \ddots & 0            \\
  \vdots       & \ddots       & \ddots       & \ddots & t^d_{\ell}            \\
  0            & \dots        & 0            & t^d_{\ell}      & t_{\ell} \\
  \end{matrix}\right)
\end{equation}
The matrices ${\cal E}(i)$ and $T_\ell$ can be diagonalized simultaneously by the same unitary transformation and the eigenvalues are 
$\epsilon(i)+t_0^d \varepsilon(k)$ and $t_{\ell}+t^d_\ell \varepsilon(k)$ respectively, where
\begin{equation}
\label{varepsilon}
\varepsilon(k)=2\cos\left(\frac{\pi k}{S+1}\right)
\end{equation}
Notice that we are considering a planar geometry, namely the coupling of the chains is open at the boundary. If we consider instead periodic boundary condition, namely the first and the last chains are coupled, the matrices Eqs.~(\ref{calES}),~(\ref{TjS}) would have had elements $t^d_0$ and $t^d_\ell$ respectively at the right-top and bottom-left corners. In that case Eq.~(\ref{varepsilon}) should be replaced by 
$\varepsilon(k)=2\cos\left(\frac{2\pi k}{S}\right)$. 
The $S$-chains model, therefore, can be decoupled to $S$ Aubry-Andr\'e chains, labeled by the index $k=1,\dots S$, 
that satisfy the following eigenvalue equations
\begin{eqnarray}
 &&\hspace{-0.7cm}\big[ E_n - \epsilon(i) - t^d_{0}\, \varepsilon(k) \big]\psi^k_{n,i} \\
\nonumber &&\phantom{----}
=\sum_{\ell}\left(t_{\ell} + t^d_{\ell}\,\varepsilon(k)
\right)
 \left( \psi^k_{n,i+\ell} + \psi^k_{n,i-\ell} \right).
\end{eqnarray}
\textcolor{black}{In the case of several coupled chains a clarification about 
the localization phase is in order. The system can be considered as 
two-dimensional with size $S\times L$, 
and since in the direction of the array of the  
chains, let us call it $y$-direction, the couplings are due to 
homogeneous hopping parameters, 
the localization of the wavefunctions induced by the quasiperiodic potential 
can occur only along the direction of the Aubry-Andr\`e chains, $x$-direction.
As a result, for finite systems, the IPR of a generic $n$-th normalized eigenstate will be limited by
\begin{equation}
\frac{1}{S L} < I_P^{(n)} \lesssim \frac{1}{S},
\end{equation}
more precisely, for open boundary conditions, $I_P^{(n)} < \frac{3}{2S}$.
}
\subsection{Nearest-neighbor hopping}
\begin{figure}[!ht]
 \centering
\includegraphics[scale=0.24]{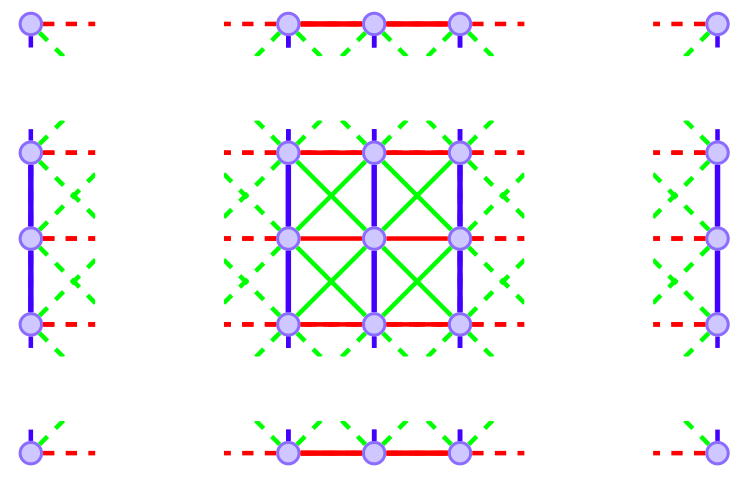}
 \caption{Many coupled chains: the red line describes the nearest-neighbor intra-chain hopping $t_{1}$, 
the blue line the transverse inter-chain hopping $t^d_{0}$, the green line nearest-neighbor inter-chain hopping $t^d_{1}$.}
 \label{fig:schemeNNS}
 \end{figure}

\begin{figure}[h!]
  \centering
  \includegraphics[scale=0.24]{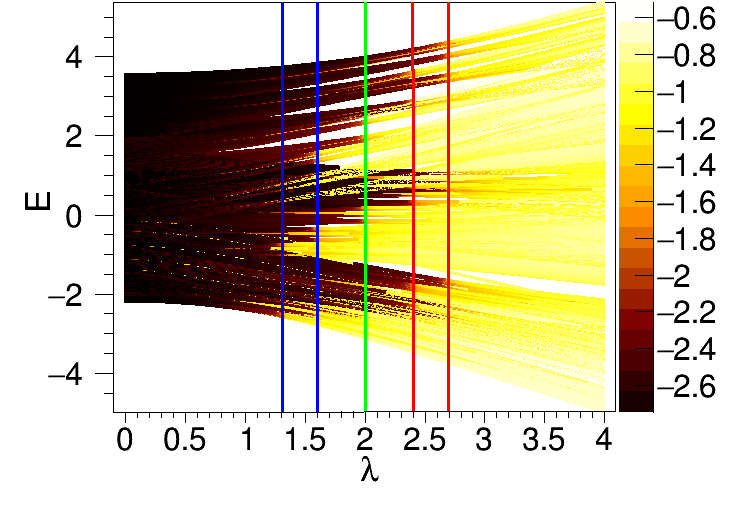}
  \caption{Inverse participation ratio (IPR), in \textcolor{black}{$10$-base} logarithmic scale, for $S=5$ coupled Aubry-And\'e chains \textcolor{black}{($L=200$)} with nearest-neighbor hopping
$t_{1} = 1$, $t^d_{1} = 0.2$, and $t^d_{0} = 0.5$. \textcolor{black}{The vertical lines are described by Eq.~(\ref{eq:lkc}), with $k=1,\dots,S$, from right (red line) to left (blue line).}}
  \label{fig:NNS5}
\end{figure}

Let us consider the nearest-neighbor hopping case ($j=i\pm 1$) depicted in Fig.~\ref{fig:schemeNNS}.  
The $S$ effective uncoupled Aubry-Andr\'e chains have the following critical potentials 
\begin{equation}
\label{eq:lkc}
 \lambda^{k}_c = 2\left(t_{1} + t^d_{1}\,\varepsilon(k)
\right)
\end{equation}
As shown in Fig.~\ref{fig:NNS5}, the critical potentials divide the phase diagram into three regions.
 \begin{figure}[tb]
  \centering
  \includegraphics[scale=0.32]{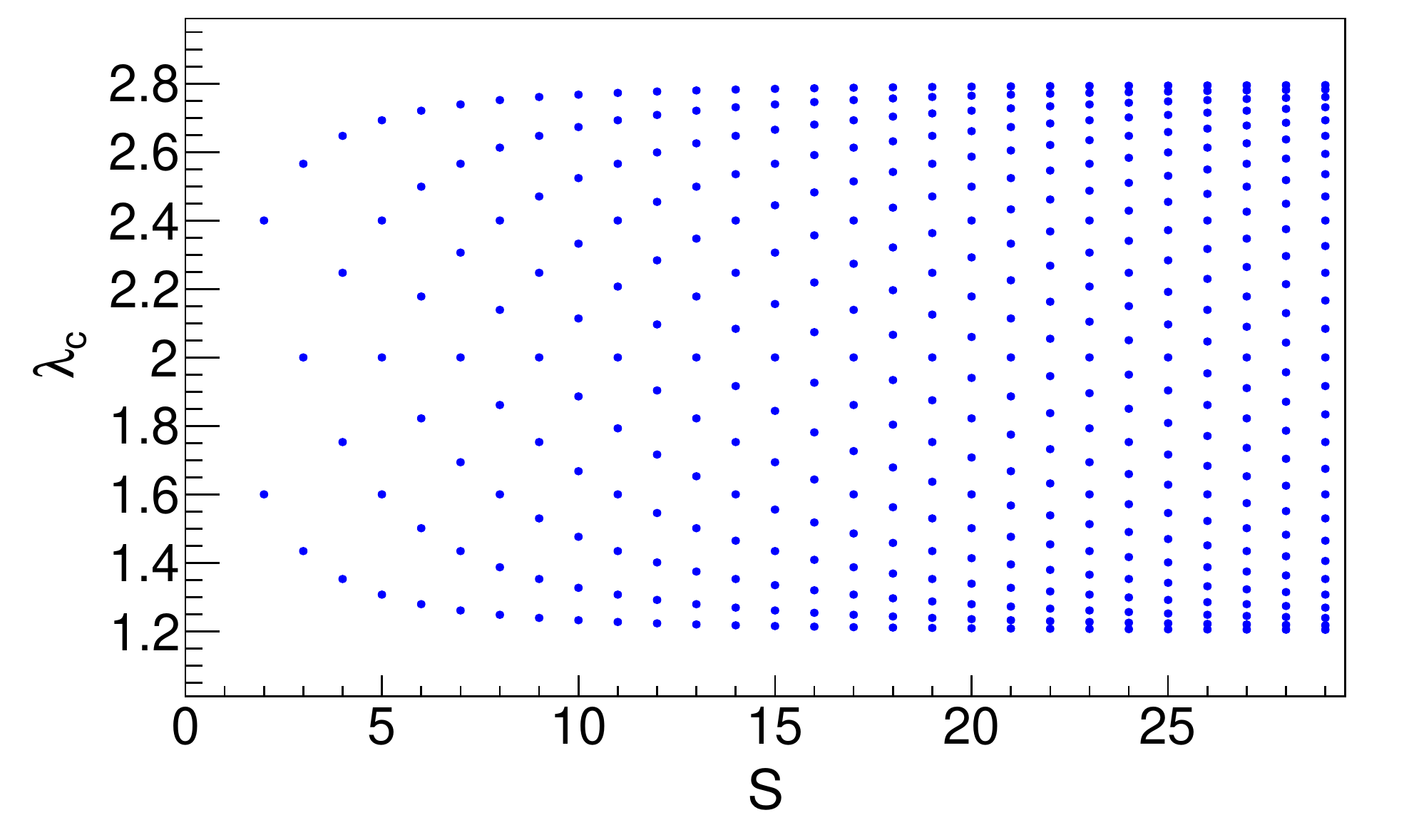}
  \caption{Critical amplitudes of the effective uncoupled Aubry-Andr\'e chains, Eq.~(\ref{eq:lkc}), for different numbers $S$ of originally 
coupled chains, for $t_{1} = 1$ and $t^d_{1} = 0.2$.}
  \label{fig:lkc}
\end{figure}
For \textcolor{black}{$\lambda < \lambda^S_{c}$} all the eigenstates are extended while for \textcolor{black}{$\lambda > \lambda^1_{c}$} are all localized.
For \textcolor{black}{$\lambda^S_{c} < \lambda < \lambda^1_{c}$}, instead, there is an intermediate region where localized and extended states coexist.
This region increases with the number of chains but is delimited by $t_{1} - 2 t^d_{1}<\lambda < t_{1} + 2 t^d_{1}$, as shown in Fig.~\ref{fig:lkc}.

\subsection{Next-nearest-neighbor hopping}
\begin{figure}[!ht]
 \centering
\includegraphics[scale=0.16]{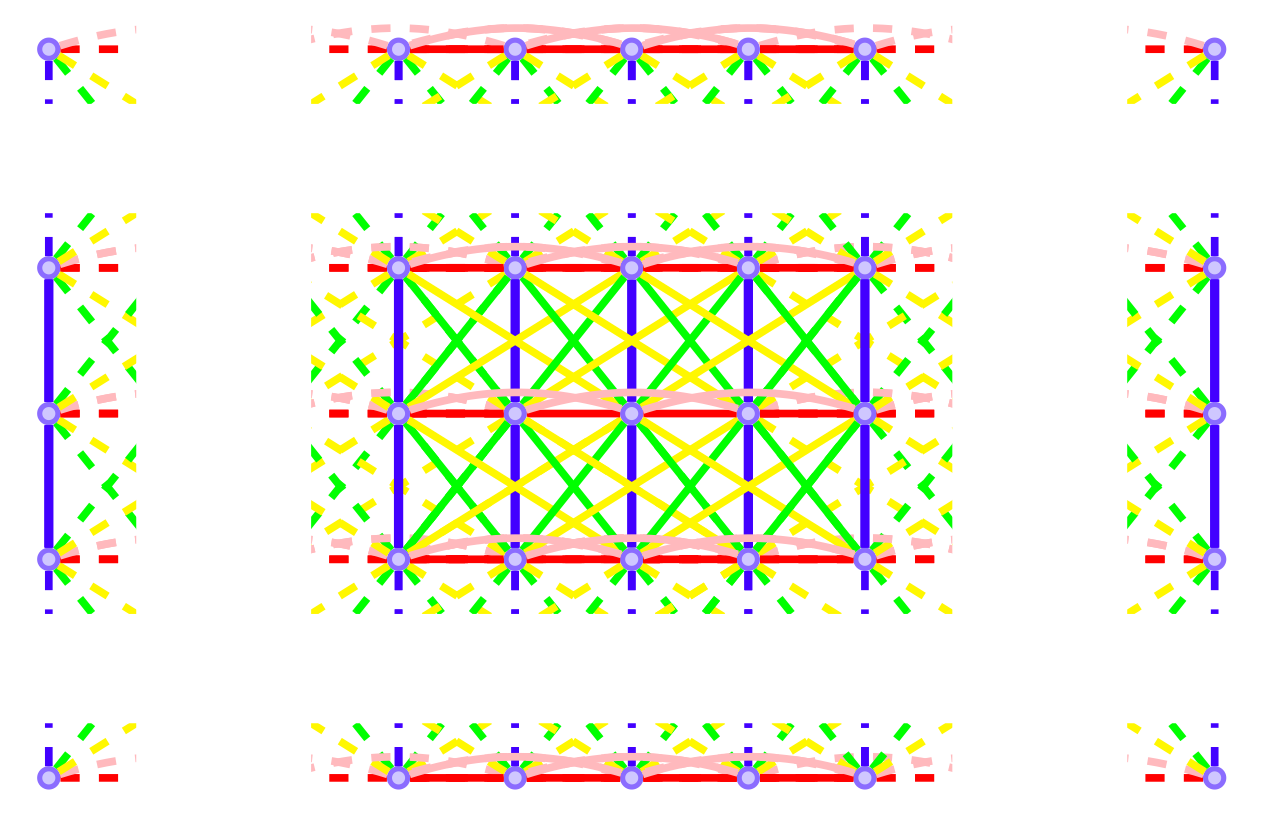}
  \caption{Many coupled chains: the red line describes the nearest-neighbor intra-chain hopping $t_{1}$,
the blue line the transverse inter-chain hopping $t^d_{0}$, the green line nearest-neighbor inter-chain hopping $t^d_{1}$, 
 the pink line the next-nearest-neighbor hopping $t_{2}$, the yellow line the inter-chain hopping $t^d_{2}$.}
 \label{fig:schemeNNNS}
 \end{figure}

\begin{figure}[ht]
  \centering
  \includegraphics[scale=0.24]{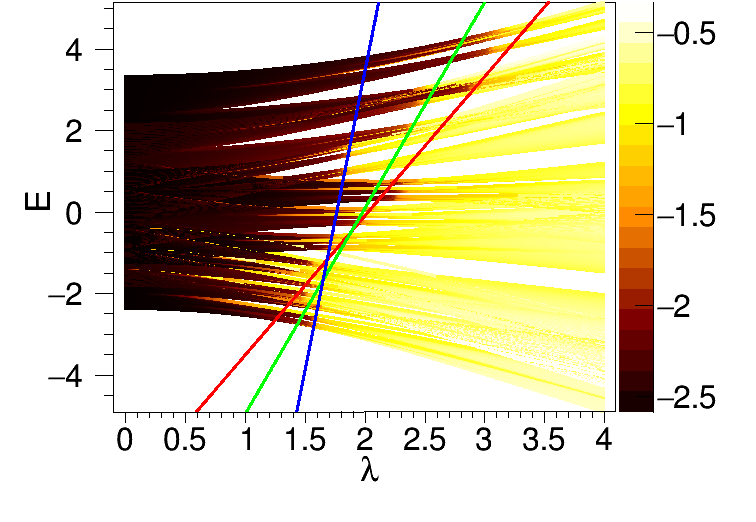}
  \qquad\qquad
  \includegraphics[scale=0.24]{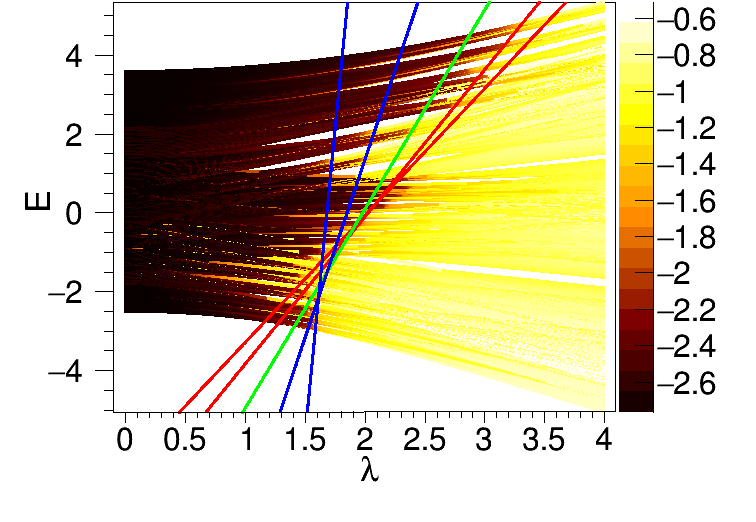}
  \caption{Inverse participation ratio (IPR), in \textcolor{black}{$10$-base} 
logarithmic scale, for $S=3$ and $S=5$ coupled Aubry-And\'e chains 
\textcolor{black}{($L=200$)} 
with next-nearest-neighbor hopping parameters $t_{1} = 1$, $t_2=t^d_{1} = 0.1$, $t^d_{2} = 0.05$ and $t^d_{0} = 0.5$. 
The straight lines are given by Eq.~(\ref{lkc_NNN}), \textcolor{black}{with $k=1,\dots,S$, from red to blue lines in a counterclockwise order.}}
  \label{fig:NLL3s200t11t1201td01td2005tds05}
\end{figure}

\begin{figure}[ht]
  \centering
  \includegraphics[scale=0.24]{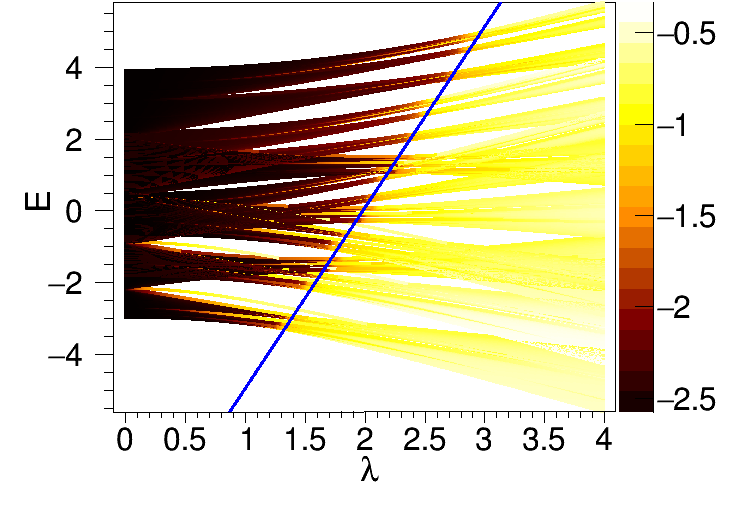}
  \caption{Inverse participation ratio (IPR), in \textcolor{black}{$10$-base} 
logarithmic scale, for $S=3$ coupled Aubry-And\'e chains 
\textcolor{black}{($L=200$)}
with next-nearest-neighbor hopping parameters $t_{1} = 1$, $t^d_{1} = t_{2} = 0.1$, $t^d_{2} = 0.01$, $t^d_{0} = 1$. 
\textcolor{black}{Since Eqs.~(\ref{cond1}), (\ref{cond2}) are fulfilled, 
so that the critical line can be written in terms of only the ratio $t_1/t_2$, 
the mobility edge, described by the blue straight line, is given 
by Eq.~(\ref{lambda_c_extended})}.
}
  \label{fig:NLL3s200t11t1201td01td2001tds1}
\end{figure}

Let us now consider next-nearest-neighbor hopping terms (Fig.~\ref{fig:schemeNNNS}), 
supposing that further terms are exponentially small. In this case, after decoupling 
the chains we get the following $S$ critical Aubry-Andr\'e amplitudes
\begin{equation}
 \lambda^{k}_c = \frac{2\left( t_{1} + t^d_{1}\,\varepsilon(k)
\right) + 2 \left( E_n - t^d_{0}\,\varepsilon(k)
\right)A_{k} }  {1 + A^{2}_{k}}
\label{lkc_NNN}
\end{equation}
where 
\begin{equation}
A_{k} = \frac{t_{2} + t^d_{2}\,\varepsilon(k)
}{t_{1} + t^d_{1}\,\varepsilon(k)}
\label{Ak}
\end{equation}
These expressions are the generalization of Eqs.~(\ref{eq:1lambdaNNN})-(\ref{A-}), already seen for two chains. 
Examples of the transitions obtained for three and five coupled Aubry-Andr\'e chains are given in 
Fig.~\ref{fig:NLL3s200t11t1201td01td2005tds05}.
If we now impose the condition $\lambda^k_{c} = \lambda^{k'}_c$, solving this equation in terms of the energy we get
\begin{eqnarray}
&&\nonumber \hspace{-0.45cm} E^{*}     = \frac{1}{(A_{k'}-A_{k})(1-A_{k'}A_{k})}
\big[
(1+A_{k'}^{2})(t^d_{1} - A_{k}t^d_{0})\varepsilon(k)
\\
&& \phantom{--} -(1+A_{k}^{2})(t^d_{1} - A_{k'}t^d_{0})\varepsilon(k')
+t_{1}(A_{k'}^{2} - A_{k}^{2})\big] 
\label{E*kk}
\end{eqnarray}
For all $k$ and $k'$ such that 
$(A_{k'}-A_{k})(1-A_{k'}A_{k})\ne 0$, we can get $\frac{S(S-1)}{2}$ different solutions $E^{*}$ 
(see for instance Fig.~\ref{fig:NLL3s200t11t1201td01td2005tds05}). 
If, otherwise, $A_{k} = A_{k'}$ or equivalently, in terms of the hopping parameters, $t_{1} t^d_{2} = t_{2}t^d_{1}$, which is the 
same condition as before, reported in Eq.~(\ref{cond1}), 
all the straight lines described by Eq.~(\ref{lkc_NNN}) have the same slope, independently from $k$, that is  
$\frac{2 t_2/t_1}{1 + \left( \frac{t_{2}}{t_{1}}\right)^{2} }$. 
If we now impose a further condition, $t^d_{0} = \frac{t^d_{1}}{t_{2}} t_{1}$, Eq.~(\ref{cond2}), 
all the parallel lines overlap each other as shown in Fig.~\ref{fig:NLL3s200t11t1201td01td2001tds1} and we get a unique mobility edge,  
\textcolor{black}{expressed again by Eq.~(\ref{lambda_c_extended})},   
dividing the extended states from the localized ones \textcolor{black}{in the $x$-direction.} 

\subsection{Long-range hopping among the chains}
A further generalization of what seen so far can be obtained allowing for longer-range coupling among the chains, 
although the physics is qualitatively the same \textcolor{black}{as that seen in the last case}. 
Imposing for simplicity periodic boundary conditions in the transverse 
direction, namely, along the array of the chains, 
assuming an exponential decay of the hopping terms, Eq.~(\ref{lkc_NNN}) should be replaced by 
\begin{equation}
\label{lamc_hop_chain}
 \lambda^{k}_c = \frac{2\left( t_{1} + {\cal P}_{1}(k)
\right) + 2 \left( E_n - {\cal P}_{0}(k)
\right)A_{k} }  {1 + A^{2}_{k}}
\end{equation}
where now $A_k$ is also replaced by 
\begin{equation}
\label{Ak_gen}
A_{k} = \frac{t_{2} + {\cal P}_{2}(k)}
{t_{1} + {\cal P}_{1}(k)}
\end{equation}
\textcolor{black}{depending on the spectrum along the transverse direction 
\begin{equation}
{\cal P}_{\ell}(k)= 2\sum_{\gamma=1}^{S-1}t^d_{\ell, \gamma}\cos(2\pi \gamma k/S).
\end{equation}}
It is worth remembering that the validity of Eq.~(\ref{lamc_hop_chain}) is based on the assumtion that $\left(t_\ell+{\cal P}_{\ell}(k)\right)$ decays exponentially with $\ell$, so that $A_k$ in Eq.~(\ref{Ak_gen}) is the exponential factor. Referring to the Hamiltonian in Eq.~(\ref{eq:ham1}), the hopping parameters are defined by
$t^d_{0,\gamma}=t^d_{0,\alpha\, \alpha\pm\gamma}$ and $t^d_{\ell,\gamma}=t^d_{i\, i\pm\ell, \alpha\,\alpha\pm\gamma}$. \textcolor{black}{Also in this case, 
imposing the condition $\lambda_c^k=\lambda_c^{k'}$ we  get the crossing 
points 
as in Eq.~(\ref{E*kk}), where the terms 
$t_\ell^d \varepsilon(k)$ are replaced by 
${\cal P}_\ell(k)$, in the hypothesis of a non-vanishing denominator. \\
If, instead we impose $A_k=A_{k'}$, 
for any $k$ and $k'$, we get the condition 
\begin{equation}
\frac{t_{1}}{t_2} = \frac{t^d_{1,\gamma}}{t^d_{2,\gamma}},\;\;\, 
\forall \,\gamma\,,
\end{equation}
as in Eq.~(\ref{cond1}), so that $A_k=t_2/t_1$ drops the $k$-dependence. 
In this condition, requiring $\lambda_c^k=\lambda_c^{k'}$, 
we obtain 
\begin{equation}
\frac{t_{1}}{t_2} = \frac{t^d_{0,\gamma}}{t^d_{1,\gamma}},\;\;\, 
\forall \, \gamma\,,
\end{equation}
as in Eq.~(\ref{cond2}), 
which brings to have a unique mobility edge described by Eq.~(\ref{lambda_c_extended}).
In conclusions, also in the most general case treated here, where all 
the several chains are coupled together, 
Eq.~(\ref{lamc_hop_chain}) can be reduce 
to Eq.~(\ref{lambda_c_extended}), the same mobility edge as that of 
a single chain with exponentially decaying hopping terms. 
}

\section{2D Aubry Andr\'e model: square lattice with quasi-periodicity in both directions}
Let us now consider a final further generalization, 
coupling (by simply transverse nearest neighbor hopping $t_0^d$) 
$S$ different Aubry-Andr\'e chains with single-site shifted energies 
between two neighboring chains. 
For $S$ and $L$ both very large we get a generalization of the Aubry-Andr\'e 
model in two dimensions (2D), 
being the quasiperiodic potential in both directions. 
The system is described by the following Hamiltonian
\begin{align}
 \ham = \sum_{\langle i, j\rangle,\alpha} t_{1} \adj{\op{c}}_{i,\alpha} \op{c}_{j,\alpha} + 
\sum_{i,\langle\alpha ,\beta\rangle} t^d_{0}\, 
\adj{\op{c}}_{i,\alpha} \op{c}_{i,\beta} 
+\sum_{i,\alpha} \epsilon(i,\alpha)
\adj{\op{c}}_{i,\alpha} \op{c}_{i,\alpha} 
 \label{eq:ham2D}
\end{align}
where we will consider $t_1=t_0^d=1$ (the same hopping in both $x$ and $y$ directions). The on-site energies are obtained by shifting the chains, namely,
\begin{equation}
\epsilon(i,\alpha)=\lambda\cos\left(2\pi \tau (i+\ell \alpha)\right).
\label{en1}
\end{equation}
In general one can use 
$\epsilon(i,\alpha)=\lambda\cos(2\pi \tau (\ell_x i+\ell_y \alpha))$, so that 
for $\ell_y=0$ and $\ell_x=1$, we recover the result for quasiperiodic potential in $x$-direction, as seen before, and for $\ell_y=1$ and $\ell_x=0$ the same but with the potential in $y$-direction. 
In these cases we have the usual critical potential, 
$\lambda_c=2t_1$ or  $\lambda_c=2t_0^d$, as in the standard Aubry-Andr\'e model, because of the perfect decoupling. We will consider also the case where $\ell$ is a random value which takes values $-1, 0, 1$, meaning that the chains are randomly shifted by at most one lattice step. \\
We will consider also a truly aperiodic 2D Aubry-Andr\'e model using the potential
\begin{equation}
\epsilon(i,\alpha)=\lambda\cos(2\pi \tau i)+\lambda\cos(2\pi p\, \alpha),
\label{en2}
\end{equation}
where $p$ is an irrational number, we will take $p=\sqrt{2}$.

\bigskip
\subsection{Two coupled chains with shift}
\begin{figure}[h!]
  \centering
  \includegraphics[scale=0.24]{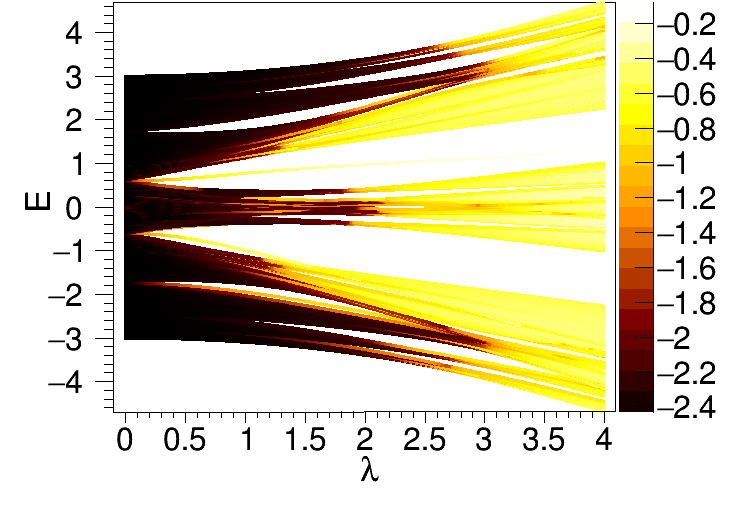}
  \caption{
  \textcolor{black}{Inverse participation ratio (IPR), in $10$-base log-scale, for two coupled Aubry-Andr\'e chains ($S=2$ and $L=200$), with single-site shifted energies, $\epsilon(i+1,1)=\epsilon(i,2)$, and 
with nearest-neighbor hopping $t_{1} = t^d_{0} = 1$.}}
  \label{fig:shift}
\end{figure}

Let us first consider two chains, which are now not identical, coupled by a simple transverse hopping $t_0$, and the intra-chain hopping parameter is only $t_1$. The energies of the first chain $\varepsilon(i,1)$ are equal to the 
energies of the second chain after a translation of $\ell$ sites, 
$\varepsilon(i,2)=\varepsilon(i+\ell,1)$ (with $\ell$ an integer number). 
We can rewrite the Hamiltonian as in 
Eq.~(\ref{Hspinor}) with  
${\cal E}(i) =\left(
\begin{matrix}
 \epsilon(i) & t^d_{0}      \\
 t^d_{0}       & \epsilon(i+\ell)
\end{matrix}\right)$
and
$T_1 =\left(
\begin{matrix}
 t_{1} & 0      \\
 0     & t_{1}
\end{matrix}\right) $.
Even if ${\cal E}$ and $T_1$ trivially commutes so that they can be 
diagonalized simultaneously as in the case of two identical chains, 
since the transformation of the fields is not global, we cannot decouple the 
systems into two uncoupled Aubry-Andr\'e chains, as done before. 
Alternatively one can show that choosing a different basis the energy term is 
not diagonalizable, see Appendix \ref{app}. 
We have, therefore, to resort to numerical exact diagonalization. 
An example of the IPR for a system of two coupled chains with $t_1=t_0^d$ and 
with shift in energies $\ell=1$ is reported in Fig.~\ref{fig:shift}. 
It is important to note that the transition between localized and extended 
states depends strongly on the energy level, in contrast to what happens 
for identical chains ($\ell=0$) where the transition occurs at 
$\lambda=2 t_1$, independently of the energy.
\subsection{Many coupled chains with shift}
\begin{figure}[h!]
  \centering
  \includegraphics[scale=0.24]{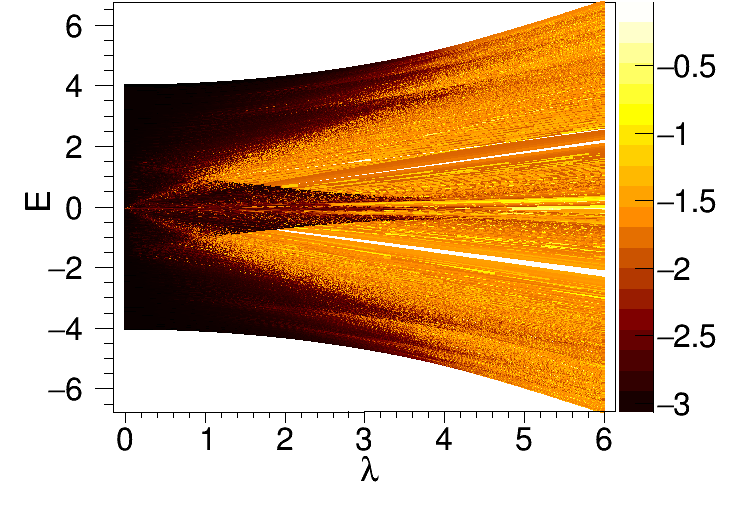}
  \caption{
  \textcolor{black}{Inverse participation ratio (IPR), in $10$-base log-scale, 
for many coupled Aubry-Andr\'e models, Eqs.~(\ref{eq:ham2D}), (\ref{en1}), with $L =S= 50$, 
and nearest-neighbor hopping $t_{1} = t_0^{d} = 1$.}}
  \label{fig:2d}
\end{figure}

We can now put together more than two chains with single-site energy shifts 
between nearest neighbor chains. When the number of chains $S$ is of the same 
order of the length of the chains $L$ we realize the generalization of the 
Aubry-Andr\'e model in two dimensions, on a square lattice. 
For $t_0^d=t_1=1$ and $S=L=50$ the results for IPR in logarithmic scale 
($\log_{10}(I_P^{(n)})$ for any energy levels $n$) 
are reported in Fig.~\ref{fig:2d}. Although the IPR is 
expected to takes value between $1/SL$ and $1$, 
we observe that, even for large potential $\lambda$, the IPR is far from $1$ 
(its log is far from $0$) meaning 
that the eigenstates are far from being strongly localized. 
This result suggests that, even for large quasidisorder potential, the
eigenstates remain somewhat extended, with values of $I_P$ of order 
of magnitude $1/L$.\\
This behavior can be explained in the thermodynamic limit, for $S,L\rightarrow \infty$, where a transverse periodicity occurs, by applying the transformation $\op{c}_{i,\alpha}=\sum_{k}e^{-\imath k (i-\alpha)}\op{c}_{i+\alpha}(k)$, 
so that the Hamiltonian, Eq.~(\ref{eq:ham2D}) with Eq.~(\ref{en1}), can be written as $\ham=\sum_{k}\ham_k$, where
\begin{equation}
 \ham_{k} = t(k)\hspace{-0.2cm}\sum_{i,\delta=\pm 1}\adj{\op{c}}_{i}(k) \op{c}_{i+\delta} (k)  
                      +\sum_i \epsilon(i)\adj{\op{c}}_{i}(k) \op{c}_{i} (k)
\end{equation}
with $\epsilon(i)=\lambda\cos(2\pi\tau i)$ and 
$t(k)=2t_1\cos(k)$, for $t_1=t_0^d$. 
As a result the system is decoupled to infinitely many 1D Aubry-Andr\'e models 
labeled by the mode numbers $k$, for which the transition occurs at 
$\lambda=2 |t(k)|$. This means that for $\lambda<4 t_1$ we have localized and extended states while for $\lambda>4t_1$ we have only localized states (as clearly shown in Fig.~\ref{fig:2d}), still in $(\op{x}-\op{y})$-direction, being the system periodic, the states remain extended. The localization is therefore only partial so that, for $L\sim S$, the IPR goes like $1/L$ for large $\lambda$ (see Fig.~\ref{fig:comp}, the red curve for the average IPR). \\
A stronger localization can be obtained by coupling chains which are randomly shifted, namely with energies as in Eq.~(\ref{en1}) but where $\ell$ takes random integer values. The corresponding IPR is more blurred than that of Fig.~\ref{fig:2d} while for large $\lambda$ it behaves like the IPR of the 2D Anderson model (see the violet dashed line in Fig.~\ref{fig:comp} which is the average IPR, where $\ell$ takes randomly the values $-1, 0, 1$, uniformly distributed). 
%
\subsection{Truly aperiodic 2D Aubry-Andr\'e model}
\begin{figure}[h!]
  \centering
  \includegraphics[scale=0.24]{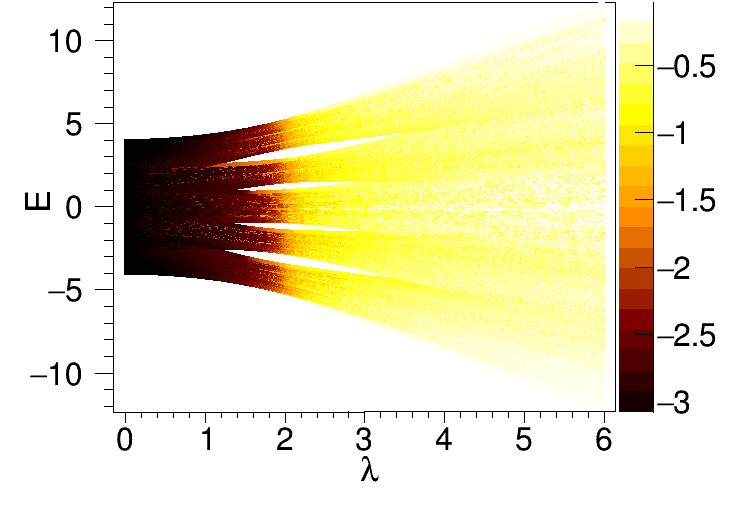}
  \caption{
  \textcolor{black}{Inverse participation ratio (IPR), in $10$-base log-scale,
for the 2D Aubry-Andr\'e, with energies given by Eq.~(\ref{en2}), with $p=\sqrt{2}$, and where $L =S= 50$, and 
nearest-neighbor hopping $t_{1} = t_{d} = 1$.}}
  \label{fig:2dAA}
\end{figure}

Finaly, let us consider a truly aperiodic 2D Aubry-Andr\'e model, with on-site energies given by Eq.~(\ref{en2}) (we used $p=\sqrt{2}$). In this case the system is not periodic in any directions. 
We calculated the eigenvalues and eigenstates numerically finding that the IPR exhibits again a sharp phase transition at 
$\lambda=2$ (in units of $t_1=t_0^d$), as one can clearly see from Fig.~\ref{fig:2dAA}. In this case the all states are localized for $\lambda>2$ and the localization is more pronunced than that of the prevous system, made by shifted chains, and that of a 2D Anderson model, with the IPR which goes rapidly to $1$. 
%
\subsection{Comparison with 2D Anderson model}
In order to clarify the results for the 2D Aubry-Andr\'e model it is useful to 
make a comparison with what one might obtain if the quasidisorder were 
replaced by a true uncorrelated disorder, as in the 2D Anderson model. 
We, therefore, solve the 
eigenproblem for an Hamiltonian as in Eq. ~(\ref{eq:ham2D}), with 
$\varepsilon(i,\alpha)$ replaced by random variables uniformly 
distributed between $-\lambda$ and $\lambda$. 
The value of $\lambda$ is, therefore, the strength of disorder. 
 \begin{figure}[t!]
  \centering
  \includegraphics[width=0.42\textwidth]{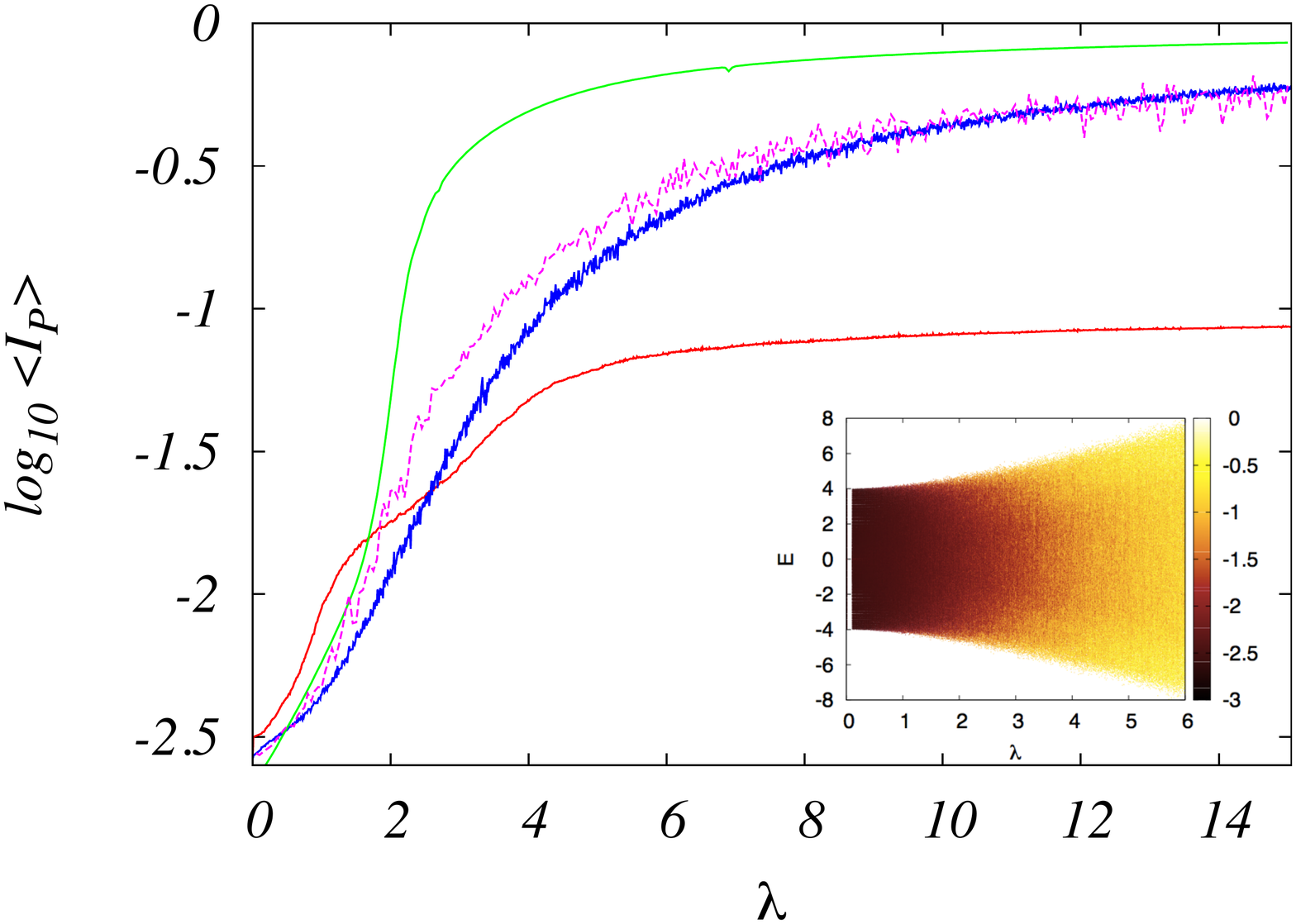}
  \caption{
  \textcolor{black}{Average inverse participation ratio, over all the eigenstates,
in $10$-base log-scale,
for the generalized 2D Aubry-Andr\'e model 
obtained by uniformly shifting the chains, Eq.~(\ref{en1}) 
(red curve), for the same Aubry-Andr\'e model but where the coupled chains are 
randomly shifted (violet dashed line), for the 2D Anderson model (blue curve), and for the truly aperiodic 2D Aubry-Andr\'e model, Eq.~(\ref{en2}) 
(green line). 
All these models are defined on a square lattice with size $L=S=30$ and 
nearest-neighbor hopping $t_{1} = t_0^{d} = 1$.
Inset: Inverse participation ratio, in $10$-base log scale, for
the 2D Anderson
model with $S=L=30$ and $t_{1} = t_0^{d} = 1$, as function of the energy levels
and $\lambda$.}}
  \label{fig:comp}
\end{figure}
In Fig.~\ref{fig:comp} we plot the average inverse participation ratio, 
$\langle I_P\rangle=\frac{1}{SL}\sum_{n=1}^{SL} I_P^{(n)}$, 
for 2D Aubry-Andr\'e model obtained by coupling many chains uniformly shifted (red curve) or randomly shifted (violet dashed line), so that the potential is given by Eq.~(\ref{en1}), or by applying an aperiodic potential like Eq.~(\ref{en2}), so that we get a truly aperiodic 2D Aubry-Andr\'e model (green line), 
in order to compare with the 2D Anderson model where the potential is an uncorrelated disorder (blue line). 
We observe that $\langle I_P\rangle$ depends weakly on $\lambda$, 
in the strong quasidisorder regime, for the system composed of shifted chains (red line). Indeed the curve is flat upon increasing 
$\lambda$, almost fixed at a value of the order of $1/L$. 
On the contrary, for the 2D Anderson model with large disorder, supposing 
$\langle I_P\rangle\sim 1/\xi^2$, where $\xi$ is the localization length, 
we find numerically that 
$\xi\sim e^{c/\lambda}$ (with $c\approx 5$ in units of $t_1=t_0^d$) in agreement with 
the theory of disorder systems in two dimensions \cite{Abrahams}. 
For uncorrelated disorder, 
therefore, the eigenstates, on average, are much more localized than those 
obtained with the Aubry-Andr\'e potential in Eq.~(\ref{en1}). 
This finding suggests  
that the connectivity in a quasidisordered network is much higher than 
that of a disordered one. 
On the contrary, if the 2D system is made by randomly coupling many Aubry-Andr\'e chains (violet dashed line) the localization effect is similar to the 2D Anderson model on average. A definitely stronger and sharper transition to localization regime is obtained by applying an aperiodic potential in both directions as described by Eq.~(\ref{en2}) (green line).

\section{Conclusions}
We studied the physics of coupled 
Aubry-Andr\'e models showing how an intermediate phase can appear, 
where localized and extended states coexist. This coexistence is actually a mixture of states which can be understood easily after decoupling the system and getting effective decoupled Aubry-Andr\'e chains with different transition points. 
\textcolor{black}
{We suggest that a weak coupling among the chains that are produced in 
the experiments can contribute to the discrepancy between the theoretical 
predictions of a vanishing intermediate phase and the 
observations of a wider regime of coexistence in the tight binding limit.}
We derive the conditions under which there is a unique well-defined mobility edge in such coupled systems 
that separates unambiguously the localized from the extended wavefunctions.
\textcolor{black}{Finally we study some localization properties in the case of a 2D Aubry-Andr\'e model obtained by coupling several chains with shifted on-site energies, finding that the extension of the wavefunctions is, on average, 
much greater than that of the states obtained solving the Anderson model. 
On the contrary, using a quasiperiodic potential in both directions as described in Eq.~(\ref{en2}) the localization regime is more pronounced and a sharp phase transition occurs.}
\section*{Acknowledgements}
LD acknowledges financial support from the BIRD2016 project of the University of Padova.

\appendix
\section{Two shifted coupled chains}
\label{app}
We can rewrite the 
Hamiltonian (\ref{eq:ham2D}), for two chains with energies 
(\ref{en1}) and $\ell=1$, similarly to  
Eq.~(\ref{Hspinor}), introducing the spinor $\hat{b}_{i}=\left(
\begin{matrix} \hat{c}_{i,1}\\ \hat{c}_{i-1,2}
\end{matrix}\right)$, also in the following way
\begin{align}
\ham = \sum_{i} \adj{\op{b}}_{i} {\cal{E}}(i) \op{b}_{i} + \sum_{i} \left(\adj{\op{b}}_{i} T^t\, \op{b}_{i+1}
+ \adj{\op{b}}_{i+1} {T}\, \op{b}_{i}\right) ,
\end{align}
with 
${\cal E}(i) =\left(
\begin{matrix}
 \epsilon(i) & 0    \\
 0       & \epsilon(i)
\end{matrix}\right)$, 
$T =\left(
\begin{matrix}
 t_{1} & 0      \\
 t^d_0     & t_{1}
\end{matrix}\right)= 
\left(
  \begin{matrix}
   1 & 0 \\
   1 & 1
  \end{matrix} 
  \right)$  and $T^t$ its transpose matrix, where we choose   $t_1=t_0^d=1$.
Writing 
\begin{equation}
\phi(i) = 
\left(
 \begin{matrix}
  \psi^{(1)}_{n,i}\\
  \psi^{(2)}_{n,i-1}
 \end{matrix}
\right)
 \end{equation}
we get the following equation
\begin{align}
 \left[ E - \epsilon(i) \right] \phi(i) = T \phi(i-1) + T^{t} \phi(i+1)
\end{align}
Let us introduce the following transformation 
$\phi(i) = \left( e^{\imath\alpha} \sigma_{y}\right)^{i} \tilde{\phi}(i)$, with $e^{\imath\alpha}$ an arbitrary phase, and $\sigma_y$ the second Pauli matrix, so that 
\begin{align}
\nonumber \left[ E - \epsilon(i) \right] \left( e^{\imath\alpha} \sigma_{y} \right)^i\tilde{\phi}(i) & = 
 T\left( e^{\imath\alpha}\sigma_{y} \right)^{i-1}\tilde{\phi}(i-1) \\
 & + T^{t}\left( e^{\imath\alpha}\sigma_{y} \right)^{i+1}\tilde{\phi}(i+1)
\end{align}
We have to distinguish two cases: when $i$ is even ($i = 2\ell$),
\begin{align}
\nonumber  \left[ E - \epsilon(2\ell) \right] \tilde{\phi}(2\ell) & = 
  T \sigma_{y} e^{-\imath\alpha}\tilde{\phi}(2\ell-1) \\
 & + T^t \sigma_{y} e^{\imath\alpha}\tilde{\phi}(2\ell+1)
\label{odd} 
\end{align}
or when $i$ is odd ($i = 2\ell+1$), 
\begin{align}
\nonumber  \left[ E - \epsilon(2\ell+1) \right] \tilde{\phi}(2\ell+1) & = 
 \sigma_{y} T e^{-\imath\alpha} \tilde{\phi}(2\ell) \\
 & + \sigma_{y} T^{t} e^{\imath\alpha}\tilde{\phi}(2\ell+2) 
\label{even}
\end{align}
Since $[T\sigma_{y}, T^{t}\sigma_{y}] =0$ and $ [\sigma_{y}T, \sigma_{y}T^{t}] = 0$, 
we can find a basis to decouple the hopping terms in both cases.  
Making the transformation 
$\tilde{\phi}(i) =P(i)\Psi(i)$, where
\begin{equation}
 P(i) =\left(
 \begin{matrix}
  1 & 1 \\
  \frac{-(- 1)^{i} - \imath \sqrt{3}}{2} & \frac{-(- 1)^{i} + \imath 
\sqrt{3}}{2}
 \end{matrix}
 \right),
\end{equation}
and calling $P_{o} = P(2\ell-1)$ and $P_{e} = P(2\ell)$, after applying $P_o^{-1}$ in Eq.~(\ref{odd}) and $P_e^{-1}$ in Eq.~(\ref{even}), we get for the both cases
\begin{equation}
  \left[ E - \epsilon(i) \right] \tilde{E}(i)\Psi(i) = 
 {D}_{L} \Psi(i-1) + {D}_{R} \Psi(i+1) 
\end{equation}
%
%
%
where
\begin{eqnarray}
\nonumber D_{L} &=& e^{-\imath\alpha}P^{-1}_{o}T \sigma_{y} P_{o} = e^{-\imath\alpha}P^{-1}_{e}\sigma_{y} T  P_{e}\\
&=&e^{-\imath\alpha}\left( \begin{matrix} \frac{-\sqrt{3} -\imath}{2} & 0 \\ 0 & \frac{\sqrt{3} - \imath}{2}\end{matrix} \right)
\end{eqnarray}
\begin{eqnarray}
\nonumber D_{R} &=& e^{\imath\alpha}P^{-1}_{o}T^{t} \sigma_{y} P_{o} =e^{\imath\alpha}P^{-1}_{e}\sigma_{y} T^{t} P_{e} \\
&=& e^{\imath\alpha}\left( \begin{matrix} \frac{-\sqrt{3} +\imath}{2} & 0 \\ 0 & \frac{\sqrt{3} + \imath}{2}\end{matrix} \right)
\end{eqnarray}
%
\begin{equation}
 \tilde{E}(i) 
=\left(
 \begin{matrix}
  1 - \frac{\imath}{\sqrt{3}}(-1)^{i} & -\frac{\imath}{\sqrt{3}}(-1)^{i} \\
  \frac{\imath}{\sqrt{3}}(-1)^{i} & 1 + \frac{\imath}{\sqrt{3}}(-1)^{i}
 \end{matrix} \right)
 \end{equation}
where $E(2\ell)=P_o^{-1}P_e$ and $E(2\ell+1)=P_e^{-1}P_o$.\\ 
$D_{L}=D_{R}^*$ and choosing $e^{-\imath\alpha} = \left(\frac{-\sqrt{3} +\imath}{2}\right)$, we get
$D_L=D_R^*=\left( \begin{matrix} 1 & 0 \\ 0 & \frac{-1 + \imath\sqrt{3}}{2}\end{matrix} \right)$. 
The matrix $\tilde{E}(i)$, instead, has only eigenvalue $1$ with multiplicity $2$, therefore not diagonalizable, and it is also 2-periodic, $\tilde{E}(i)=\tilde{E}(i+1)^*=\tilde{E}(i+2)$.

\end{document}